\documentclass{emulateapj}
\usepackage{graphicx}
\usepackage{natbib}
\usepackage{amsmath}
\usepackage{ mathrsfs }
\usepackage{subfigure}
\usepackage{epstopdf}
\usepackage{tipa}

\long\def\symbolfootnote[#1]#2{\begingroup%
\def\thefootnote{\fnsymbol{footnote}}\footnote[#1]{#2}\endgroup}

\newcommand{\beq}{\begin{equation}}
\newcommand{\eeq}{\end{equation}}
\newcommand{\bea}{\begin{eqnarray}}
\newcommand{\eea}{\end{eqnarray}}
\newcommand{\bff}{}

\newcommand{\krho}{{k_\rho}}
\newcommand{\sigmaNTc}{{\sigma_{{\rm NT},c}}}
\newcommand{\sigmac}{{\sigma_c}}
\newcommand{\Mdot}{{\dot{M}}}
\newcommand{\Mdotc}{{\dot{M}_c}}
\newcommand{\Mc}{{M_c}}
\newcommand{\Rc}{{R_c}}
\newcommand{\Mdotin} {{\dot{M}_{\rm in}}}
\newcommand{\Mdotstar}{{\dot{M}_*}}
\newcommand{\Mstar}{{{M}_*}}

\newcommand{\Mdotej}{{\dot{M}_{\rm ej} }}

\newcommand{\alphavir}{{\alpha_c}}
\newcommand{\alphares}{{\alpha_{\rm res}}}
\newcommand{\etaM}{{\eta_M}}

\newcommand{\epsin}{{\epsilon_{\rm in} }}
\newcommand{\epsinbar}{{\bar\epsilon_{\rm in} }}
\newcommand{\epsstar}{{\varepsilon_*}}
\newcommand{\xic}{{\xi_c}}
\newcommand{\xistar}{{\xi_*}}
\newcommand{\SFRff}{{{\rm SFR}_{\rm ff}}}
\newcommand{\tff}{{t_{\rm ff}}}
\newcommand{\tffc}{{t_{{\rm ff},c}}}
\newcommand{\cs}{c_{s,c}}

\newcommand{\Sigmac}{{\bar\Sigma_c}}
\newcommand{\mdotstar}{{\dot m}_*} 

\newcommand{\mbarstar}{\bar m_*}

\newcommand{\vchar}{v_{{\rm ch},w}}

\newcommand{\vescc}{v_{{\rm esc},c}}
\newcommand{\kps}{{\rm km/s}}

\newcommand{\aext}{a_{\rm ext}} 
\newcommand{\phiacc}{\varphi_{\rm acc}}

\newcommand{\lambdafil}{\lambda_{\rm fil}}
\newcommand{\Lambdafil}{\Lambda_{\rm fil}}
\newcommand{\csfil}{\sigma_{{\rm fil}}}
\newcommand{\Nfil}{N_{\rm fil}}

\newcommand{\Min}{M_{\rm in}}
 
\newcommand{\Sigmacrit}{\Sigma_{\rm crit}}

\newcommand{\ftrapr}{f_{\rm trap,\,IR}}

\newcommand{\taudstar}{\tau_{d*,i}}
\newcommand{\sigmadstar}{\sigma_{d*}}
\newcommand{\taudstarmax}{\tau_{d,{\rm max}}}
\newcommand{\taudSt}{\tau_{d*,{\rm St}}}
\newcommand{\sigmatau}{\sigma_{\tau}}
\newcommand{\nbari}{\bar n_i} 
\newcommand{\FII}{F_{\rm II}}
\newcommand{\phiIR}{\phi_{\rm IR}}
\newcommand{\varphiHMSF}{\varphi_{\rm HM}}

\begin{document}
\submitted{}

\title{Star cluster formation with stellar feedback and large-scale inflow} 
\shorttitle{Star cluster formation}
\shortauthors{Matzner and Jumper}

\author{Christopher D. Matzner \& Peter H. Jumper}
\affil{Department of Astronomy and Astrophysics, University of Toronto, 50 St. George Street, Toronto, Ontario, M5S 3H4, Canada}
\accepted{by ApJ Nov 10, 2015}
\email{matzner@astro.utoronto.ca}

\begin{abstract}
{\bff During star cluster formation, ongoing mass accretion is resisted by stellar feedback in the form of protostellar outflows from the low-mass stars and photo-ionization and radiation pressure feedback from the massive stars. }
We model the evolution of  cluster-forming regions during a phase in which both 
{\bff accretion and feedback} 
are present, and use these models to investigate how star cluster formation might terminate.  
Protostellar outflows are the strongest form of feedback in low-mass regions, but these cannot stop cluster formation if matter continues to flow in.  In more massive clusters, radiation pressure and photo-ionization rapidly clear the cluster-forming gas when its column density is too small.    We assess the rates of dynamical mass ejection and of evaporation, while accounting for the important effect of dust opacity on photo-ionization.   Our models are consistent with the census of protostellar outflows in NGC~1333 and Serpens South, and with the dust temperatures observed in regions of massive star formation.  Comparing observations of massive cluster-forming regions against our model parameter space, and against our expectations for accretion-driven evolution, we infer that massive-star feedback is a likely cause of gas disruption in regions with velocity dispersions less than a few kilometers per second, but that more massive and more turbulent regions are {\bff too strongly bound for stellar feedback to be disruptive.} 	
\end{abstract}

\section{Introduction}\label{S:Intro}

Star formation is a highly clustered and correlated phenomenon as a consequence of the clumpy nature of massive, turbulent molecular clouds, and because only the densest and most shielded regions within these clouds readily collapse to form stars.  For the birth of a star cluster within one of the molecular clumps, there are two major implications: first, that the large-scale flows which assembled the clump can  continue to rain down upon it as the star cluster is born; and second, that individual protostars are close enough in space and time to affect one another via jets, winds, radiation, and potentially supernovae.   These two effects -- input and output, or accretion and feedback -- are both capable of driving turbulent motions within the star-forming medium, affecting its dynamics and therefore the rate and nature of star formation, and adding or subtracting mass to the forming cluster.  {\bff  The purpose of this paper is to understand the competing influences of accretion and stellar feedback during a cluster's formation. }

 We are motivated by several persistent questions.   What is the dominant mode by which matter is assembled: monolithic collapse, colliding flows, or continued accretion?  What is the role of inflowing matter in the dynamical evolution of a growing cluster?  How important are the effects of stellar feedback -- in general, or for specific observed regions?  And, what ultimately ends star cluster formation: a limited supply of bound matter, perhaps, or disruption by stars?   

We  begin by considering observational results on the structure and dynamics of molecular clouds, which we use to infer how matter accumulates and accretes. The gross properties of a forming cluster are often controlled by {\bff accretion}, at least until stellar feedback becomes strong.   Turning to stellar feedback, we evaluate the effects of protostellar outflows, first with a simple comparison of forces and then by applying analytical results from  \citet{2000ApJ...545..364M} and \citet{2007ApJ...659.1394M} to an accreting molecular clump.    We estimate the direct and indirect forces of radiation pressure and of photo-ionized gas pressure, and comment on the strength of stellar wind pressure, in order to assess the feedback from massive stars.  After mapping feedback regimes, we compare against individual low-mass regions NGC~1333 and Serpens South, as well as surveys of massive cluster-forming regions. 
{\bff Finally} we draw conclusions about when and whether stellar feedback terminates star cluster formation. 

\section{Mode of mass accumulation}\label{S:Accumulation} 

The manner by which matter is gathered and enters the region of cluster formation is an important factor shaping the cluster's final properties.  Since star cluster formation is intermediate in mass and duration between the formation of giant molecular clouds and the formation of individual stars, the scenarios considered on larger and smaller scales have also been adopted for star cluster formation.  These include: collapse as a result of colliding flows \citep{1996ApJ...473..881V}, Bondi-Hoyle  \citep{2011ApJ...735...25N} or Bondi   \citep{2012ApJ...746...75M} accretion,  inside-out collapse of a turbulent virialized structure \citep{1992ApJ...396..631M,1997ApJ...476..750M,2003ApJ...585..850M}, and the gravitational infall of an initial, possibly filamentary density distribution \citep[e.g.,][]{2011ApJ...740...88P}.    While there is some overlap, these differ as to whether mass inflow overlaps the period of star formation, as to whether the in-flowing gas is bound to the cluster-forming region before being incorporated, and also with respect to the time history and detailed properties (such as density distribution, angular momentum, and magnetization) of the inflow. 

For a couple reasons we believe that observations of the molecular cloud environments for stellar cluster formation favor the latter options, i.e. inside-out collapse or infall from an initially filamentary structure, at least in the current Milky Way.

First, molecular clouds are highly filamentary.  Filaments are a prominent feature of observations of molecular clouds \citep[e.g.][]{1979ApJS...41...87S,1987ApJ...312L..45B,1989ApJ...338..902L,1999ApJ...510L..49J,2010A&A...518L.102A,2012A&A...540L..11S,2014A&A...561A..83P}.
Moreover this structure is directly relevant to the formation of star clusters. 
As \citet{2009ApJ...700.1609M} has stressed, filaments are nearly ubiquitous in the regions surrounding active star cluster formation; indeed star cluster formation is often observed at the junctions of molecular filaments \citep[for instance, in the Rosette molecular cloud;][]{2012A&A...540L..11S}.  In some cases, such as Serpens South, accretion along filaments has been inferred from molecular line kinematics \citep[e.g.,][]{2013ApJ...766..115K} or via chemical signatures \citep{2013MNRAS.436.1513F}.   
\cite{2014A&A...565A.101T}, who map N$_2$H$^+$ toward 17 {\em Herschel} filaments, find several examples of filamentary accretion flows onto clumps.  Filamentary accretion is also a prominent feature in numerical simulations of massive star formation \citep[e.g.,][]{2006MNRAS.373.1091B}, which is analogous to star cluster formation. 

Second, there exists a clear trend for the most massive clumps within molecular clouds to exhibit the lowest levels of turbulent support against gravity.  \citet{1992ApJ...395..140B} found that only the most massive regions within several molecular clouds (which are also the densest and exhibit the highest column density) had sufficiently small line widths to be considered strongly self-gravitating.    \citeauthor{1992ApJ...395..140B}  define the virial parameter $\alpha = {5\sigma(r)^2 r / G M(r)} $
to compare turbulence against gravity.   (Here $\sigma(R)$ is the one-dimensional velcocity dispersion of a region of radius $r$ and mass $M(r)$).  Reviewing a number of more recent works and  using more robust dust-derived masses, \citet{2013ApJ...779..185K} verify this trend and show that it extends to remarkably low values of $\alpha$, at least for condensations of quiescent cold gas and dust without prominent signs of star formation.  As \citeauthor{2013ApJ...779..185K} argue, this implies that the initial conditions for massive star formation are either in a state of imminent collapse or suspended by strong magnetic fields.  These conclusions {\bff apply equally well} to the initial conditions for the creation of a star cluster.  

Motivated by these points we shall concentrate on the continuous infall of gravitationally-bound matter, which is organized in a filamentary fashion toward the site of star cluster formation.   Several implications are apparent: 

\noindent{\bff --}  {\em Infall duration:} The infall of a mass reservoir $\Min(r)$, found within radius $r$ of the cluster formation site at $t=0$, will last for roughly the initial free-fall time $\tff(\Min) = [\pi^2 r^3/(8 G \Min)]^{1/2}$ (in the monopole approximation).   This duration is intermediate between the inside-out collapse of an initially hydrostatic state, for which infall takes a couple $\tff(\Min)$ because of the forces which balanced gravity in the initial state, and the rapid formation of a cluster by colliding flows of unbound gas. 

\noindent{\bff --} {\em Dynamical age of cluster formation:} Because matter accumulates in a dense central cluster-forming region as matter falls in, the free-fall time on the cluster scale should be significantly shorter than that of the reservoir.  Therefore, a proto-cluster's formation extends for several dynamical times of the its parent clump ({\bff e.g., 6.2 clump free-fall times in the fiducial model of \S~\ref{S:Params} [eq.~\ref{eq:times}]}), and one should consider the physical state of the cluster-forming region as this happens.   {\bff In particular, survival for multiple free-fall times strongly favors virial levels of clump turbulence despite sub-virial initial conditions, and this is a key feature of the model we develop below.} 

\noindent{\bff -- }{\em Specific energy of infall:}  Unless the initial conditions are highly magnetized, low values of the initial virial parameter imply that the energy per particle of the inflowing matter is close to its initial gravitational potential.  If the radius of the cluster-forming region is small compared to the initial  radius, then the inflow speed will be close to the escape velocity.  

\noindent{\bff --} {\em Time profile of infall:} The filamentary nature of the initial state implies that the mass inflow rate is reasonably constant -- neither rapidly increasing nor rapidly decreasing -- as the cluster gains most of its mass.  We illustrate this below in \S~\ref{SS:Filaments} with a simple model {\bff that displays} a constant rate of accretion.  Nevertheless, a decline in the inflow rate or nature of the inflow remains one possible cause for the end of cluster formation. 

{\bff Given these points, we favor the following model for star cluster formation.  The initial conditions correspond to elongated or filamentary molecular concentrations with low virial parameters.  While magnetic fields may support these structures locally, magnetic support along their long axes is unlikely (albeit not impossible; \citealt{1997ApJ...475..237L}).  Collapse therefore proceeds in the `rapid and violent', nearly free-fall manner envisioned by  \citet{2013ApJ...779..185K}.\footnote{\bff Energy conservation during free-fall collapse implies virial parameters of order unity (\citealt{1981MNRAS.194..809L}; \citealt{2006MNRAS.372..443B}; \citealt{2013ApJ...779..185K}).  However during {\em filamentary} collapse, infall will be visible primarily as a velocity gradient rather than a line-of-sight velocity dispersion.  The virial parameter of the collapsing reservoir will therefore remain less than unity unless $\sigma(r)$ is defined in a way that accounts for velocity gradients.}  A stellar cluster-forming clump accumulates at the centre of this collapse, undergoes star formation, and evolves under the combined effects of accretion and stellar feedback (Figure~\ref{Fig:Schematic}).  Because it gains mass over several internal free-fall times, we expect the clump to be virialized, with virial parameter $\alpha_c$ of order unity.   }

{\bff In this scenario, an important factor in a clump's evolution is its accretion parameter
\begin{equation} \label{etaM}
\etaM = {t\,\Mdotin(t) \over \Min(t)},
\end{equation} 
which compares the current rate of accretion to its historical average.}  This is fixed in the case where the mass and inflow rate are power laws of time, $\Min \propto t^\etaM$ and $\Mdotin\propto t^{\etaM-1}$.  So long as each mass shell enters the cluster in a time $t(\Min)$ which is proportional to  $\tff(\Min)$, any mass distribution with $\Min\propto r^{3-\krho}$ collapses with $\etaM = 6/k_\rho -2$.  This might represent a spherical condensation with density profile $\rho(r) \propto r^{-k_\rho}$ \citep[e.g.,][]{1997ApJ...476..750M,2003ApJ...585..850M}, but it can also capture the filamentary infall scenario of \S~\ref{SS:Filaments}.   Initial conditions with $\krho<2$ lead to accelerating accretion ($\etaM>1$), and those with a marked increase in column density toward the center, i.e. $\krho>1$, have $\etaM < 4$.  

\subsection{Filamentary infall: a simple model} \label{SS:Filaments}

If unstable filamentary structures are indeed the typical initial conditions for star cluster formation, then it is reasonable to introduce a particularly simple model of filamentary infall in which several filamentary structures emanate away from the cluster formation site at $t=0$.   To make the model specific (at some cost in realism), we neglect the motions of gas which created these filaments and arranged them in this way, assuming matter reaches the clump exclusively via filaments, i.e., without any additional accretion of non-filamentary material.     If there are filaments stretching away radially in each of $\Nfil$ directions, and the average mass per unit length along each is $\lambdafil$ at $t=0$, then the reservoir mass is $\Min(r)= \Nfil \lambdafil r$, the monopole free-fall time is $\tff(r) = [\pi^2/(8 \Nfil G \lambdafil)]^{1/2} r$, and the characteristic infall rate $\Mdotin \simeq \Min/\tff$ is constant ($\etaM=1$) with the value
\begin{equation}\label{eq:Mdotin-fromlambda}
\Mdotin \simeq \left({8\over\pi^2 }G \lambdafil^3 \Nfil^3\right)^{1/2}. 
\end{equation} 
The monopole approximation is not appropriate for $\Nfil =2$, which could describe an infinite filament with no acceleration toward $r=0$.  For $\Nfil =1$ the cluster forms at the end of a lone filament, not near the center of mass; see \citet{2011ApJ...740...88P} and \citet{2012ApJ...756..145P} for further discussion.  The monopole approximation is reasonably accurate, however,  for $\Nfil \geq3$.  

It is useful to re-express equation (\ref{eq:Mdotin-fromlambda}) in terms of the velocity dispersion within the filament. 
An infinite, axisymmetric, isothermal filament of sound speed $c_s$, supported by nothing but gas pressure, has a critical mass per unit length $2c_s^2/G$ \citep{1963AcA....13...30S,1964ApJ...140.1056O} below which it must be confined by external pressure, and above which homologous collapse ensues until a change in the equation of state causes it to fragment \citep{1992ApJ...388..392I}.   Normalizing  to the critical value for the total velocity dispersion $\csfil$, the mass per length along the filament axis is 
\begin{equation}
\lambdafil = 2\Lambdafil {\csfil^2\over G}.
\end{equation}
 The criticality parameter $\Lambdafil$ has a maximum of about unity if magnetic forces are negligible. The mass per unit radius is greater than the mass per unit length by a factor $1/\cos(\theta)$ if filaments deviate from the radial direction by an angle $\theta$, and this factor should be absorbed into $ \Lambdafil\Nfil$.   Combining this definition with equation (\ref{eq:Mdotin-fromlambda}), 
\begin{eqnarray} \label{eq:Mdot_in_Filamnetary_Infall}
\Mdotin &\simeq& \frac{8}{\pi} \left(\Lambdafil\Nfil\right)^{3/2} {\csfil^3\over G}  \\ 
&=&  1050 \left(\Lambdafil \Nfil\over 4\right)^{3/2} \left(\csfil\over 0.6\,\kps\right)^3 \,M_\odot\,{\rm Myr}^{-1}.  \nonumber
\end{eqnarray} 

Compared to the inside-out collapse of an initially static singular isothermal sphere of sound speed $\csfil$  \citep{1977ApJ...214..488S}, the infall rate is higher by the factor $2.6 (\Lambdafil\Nfil)^{3/2}$.  This, along with the fact that filaments display varying degrees of non-thermal support, allows $\Mdotin$ to reach the large values required to build a massive star cluster in about a million years, even within this very restricted set of assumptions.  Any initial inward motion will increase the accretion rate relative to this estimate.  

In extreme cases, the required accretion rate can be very high (e.g. $\sim 1\,M_\odot/$yr to build a massive globular cluster), requiring $\csfil\sim 6$\,km\,s$^{-1}$ in equation (\ref{eq:Mdot_in_Filamnetary_Infall}).  This can lead to collapse and fragmentation of the filaments before they can accrete (despite magnetic support; see \citealt{2014MNRAS.443..230H}).

\begin{deluxetable*}{ccl}
\tabletypesize{\tiny} 
\tablecaption{Variables and model parameters} 
\tablehead{ \colhead{Variable}  &   \colhead{Fidicual value}   & \colhead{Definition} } 
\startdata
\sidehead{\underline{Reservoir properties (\S~\ref{S:Accumulation})  }}
$M_{\rm in}(r)$  & \nodata &  Initial reservoir mass within $r$ \\
$t_{\rm ff}(r)$ & \nodata &  Free-fall time at $r$, monopole approximation \\
$ k_{\rho} $ & \nodata &  Effective density index, $M_{\rm in}(r)\propto r^{3-k_\rho}$     \\
$\eta_M$ & \nodata &  Accretion rate parameter: $M_{\rm in}(t)\propto t^{\eta_M}$ \\
\sidehead{\underline{Filamentary infall model (\S~\ref{SS:Filaments}) } }
$\lambda_{\rm fil};  N_{\rm fil}  $ & \nodata &   Mass per unit length; number    \\
$\Lambda_{\rm fil} $ & 1 &    Criticality parameter, $G\lambda_{\rm fil}/\sigma_{\rm fil}^2$ \\
\sidehead{\underline{Clump properties (\S \ref{S:Params}) } }
$ \epsilon_{\rm in} $ & \nodata &  Accretion efficiency, $\dot M_{\rm cl}/\dot M_{\rm in}$  \\
$ f_g $ & \nodata &  Gas fraction, $M_g/M_c$     \\
$\alpha_c$ & 1.6 &  Virial parameter, $5\sigma_c^2 R_c/(G M_c) $  \\
$ \xi_c $ & \nodata &   Turbulence-to-accretion parameter, $\sigma_c^3/(G \dot M_{\rm in})$    \\
$\varphi_{\rm acc}$ & 0.8 &  Efficiency of accretion-driven turbulence \\ 
$ \epsilon_* \xi_*$ & 0.5 &    Dimensionless stellar accretion rate, $G\dot m_*/c_{s,c}^3$   \\
$ {\rm SFR}_{\rm ff} $ &  0.03 &   Star formation efficiency per free-fall time   \\
$\sigma_c;  c_{s,c}; t; t_{{\rm ff},c}$ & \nodata &  1-D velocity dispersion; isoth.\ sound speed; age; free-fall time  \\ 
$ R_c; M_c; \bar \Sigma_c$; $M_{\rm ej}$ & \nodata &  Radius; mass; mean surface density; ejected mass, $M_{\rm in}-M_c$   \\
\sidehead{\underline{Stellar feedback (\S~\ref{S:Feedback})} }
$v_{{\rm ch}, w}$ & 30 km\,s$^{-1}$ &  Protostellar wind momentum per unit mass   \\
$\bar m_*$ &  $0.2 M_\odot$  &  Mean stellar mass \\ 
$Y$ & \nodata &   10.1 SFR$_{\rm ff}/(\alpha_c^{3/2} \xi_c)$  \\ 
$q_*$ & 1 &  d\,$\ln M_*$/d\,$\ln M_c$  \\
$\bar \kappa; \bar \kappa_R;\bar  \kappa_{\rm Pl}; \bar \kappa_{dd}$ & \nodata &  Mean opacity: flux-averaged; Rosseland; Planck; to dust emission \\ 
$L_*$; $S_*$; $(L/M)_*$; $\psi_{47}$  & \nodata &  Stars: luminosity, ionizing output; light-to-mass ratio;  $L_*/(47\,{\rm eV} S_*)$ \\ 
$F_{\rm grav}$; $F_{\rm rad, (dir, ind)}$; $F_{\rm II}$; $F_w$; $F_{\rm out}$ 
      & \nodata &   Forces:  gravity; (direct, indirect) radiation; from H\,II; wind; net outward  \\ 
 ${\cal F}_{M, A}(\Sigma/\bar \Sigma_c) $ & \nodata &  Fraction of clump (mass, area) below $\Sigma$, given $\bar \Sigma_c$   \\ 
$\Gamma; \Gamma_{\rm thin}; \Gamma_{\rm thick}$ & \nodata &  Eddington parameter of reprocessed starlight; thin limit; thick limit  \\
$\Sigma_{\rm crit}$; $\Sigma_{\rm crit, tot}$ & \nodata &  Critical $\bar \Sigma_c$ for disruption by direct radiation; for net outward force  \\ 
$1 + \phi_{\rm IR}$ & 2 &  Direct force enhancement due to innermost re-emission  \\
$\phi_w$ & \nodata &  Wind force factor, $F_w c/L_*$  \\ 
$\sigma_{d*,-21}$; $T_{i,4}$  & 1; 1 &  H\,II: dust opacity ($10^{-21}$\,cm$^{2}$/H) to starlight; temperature ($10^4$\,K)  \\  
$\tau_{d*,i}$; $\tau_{d*,{\rm St}}$; $\tau_{d,{\rm max}}$ & \nodata &  H\,II dust optical depth to ion.\ light; Str\"omgren value; maximum \\   
$\sigma_\tau$ &2.0 &   Critical $\sigma_c$ for $\tau_{d*,{\rm St}}=1$ \\ 
$\sigma_F$ & \nodata & See eqs.\ (\ref{eq:F_i}) and (\ref{eq:sigma_F})  \\ 
$\dot M_{\rm out,\, dyn}; \dot M_{\rm out,\, evap}$ & \nodata &  $\dot M_{\rm ej}$ from $F_{\rm out}$ due to massive stars; from photo-evaporation \\ 
$t_{\rm ff,c}'$, $F'_{\rm grav}$, etc. & \nodata & Quantities modified by $F_{\rm rad, ind}$  via $G\rightarrow G'=G/(1-\Gamma)$
\enddata
\label{Table:Definitions}
\end{deluxetable*}

\section{Clump evolution under rapid accretion} \label{S:Params} 

Having identified our preferred scenario for the accumulation and continued accretion of matter onto the site of star cluster formation, we now turn to the dynamics of the cluster-formation process itself.  We adopt the idealization that there exists a cluster-forming `clump' which serves as a mass reservoir for star formation, and which is dynamically distinct from the matter falling into it.  {\bff Our distinction between  the clump and its accretion flow is corroborated by the recent detection of two power-laws in the column density distributions of molecular clouds \citep{2015arXiv150708869S}, in which the lower-column component corresponds to filamentary features, while the high-column component is concentrated in knots at the intersections of filaments. }

Despite the filamentary and clumpy nature of accretion, we assume for simplicity that the clump properties can be described with a total (gas and star) mass $\Mc$, a radius $\Rc$, and one-dimensional velocity dispersion $\sigmac$ (which can be decomposed into thermal and non-thermal components: $\sigmac^2 = c_{s,c}^2 + \sigmaNTc^2$).   
\begin{figure}
\includegraphics[scale=0.3]{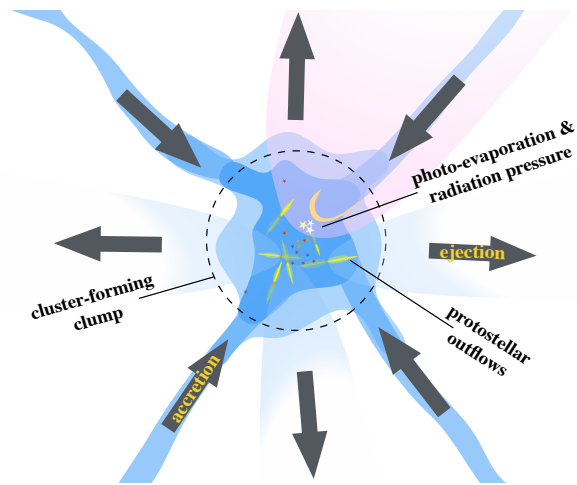}
\caption{\bff Schematic of the star cluster formation scenario.  Inflow, lasting the free-fall time of the filamentary mass reservoir, leads to the accumulation of matter over several free-fall times of a gaseous, star-forming clump, while protostellar outflows and the massive-star feedback (photo-ionization and radiation pressure) tend to eject matter.  Both feedback and inflow  drive turbulent motions.  The clump's evolution, and that of the star cluster, are determined by a competition between these effects.}
\label{Fig:Schematic}
\end{figure}
{\bff We depict this scenario in Figure \ref{Fig:Schematic}.  } 
It is worthwhile to define several additional parameters before discussing the clump's evolution. 

At any time a fraction $f_g$ of the mass is in gas and $1-f_g$ is in stars, so the gas and star masses are $M_g = f_g \Mc$ and $M_* = (1-f_g)\Mc$.   These quantities evolve as matter falls into the clump,  and collapses into stars or is blown away.  
If we normalize the rate at which gas is converted into stars to the core's free fall rate $\tffc^{-1} = [8G\Mc/(\pi^2 \Rc^3)]^{1/2}$ by the factor $\SFRff$, then 
\begin{equation} 
\Mdotstar = \SFRff {f_g \Mc\over \tffc}. 
\end{equation} 
Because mass can be lost from the clump as well as added, the clump mass can grow more slowly than mass falls in: 
\begin{equation} 
\Mdotc = \epsin \Mdotin 
\end{equation}
where $\epsin\leq1$.   Later on we will calculate $\epsin$ using a model for mass ejection due to stellar feedback. 

How might a clump evolve under the influence of accretion and feedback?  A clue to the answer lies in the fact that we introduced very few dimensional scales when describing the clump's growth.   If the  star formation and stellar feedback do not rapidly change the clump properties, and the infall history is steady ($\etaM$ does not change rapidly) then we may expect the dimensionless parameters describing clump evolution to be constant, or at least slowly varying.   This is reasonable insofar as  $\SFRff$ is reasonably constant (as in the \citealt{1989ApJ...345..782M} and \citealt{2005ApJ...630..250K} theories, for instance) and when $\epsin$ is not varying rapidly. 

Because the clump persists and grows for multiple free-fall times, it must maintain a state close to virial equilibrium, and therefore its virial parameter 
\begin{equation} \label{eq:virial}
\alphavir = {5 \sigmac^2 \Rc\over G \Mc}
\end{equation}
should remain reasonably constant.   Indeed the molecular clumps which host  star cluster formation are observed to contain turbulence comparable to the level required by virial balance.  In the \citet{2003ApJS..149..375S} study of cluster-forming CS clumps traced by water masers, the highest-quality subsample has a median $\alphavir$ of 1.6.  {\bff (We adopt this as our fiducial value for the models developed below.)}  In a theoretical study of accreting molecular clouds, \citet{2011ApJ...738..101G} found that the ram pressure of infall and outflow, both of which have a compressive effect, lead to cloud virial parameters in the range 1 to 3.  Compared with the low virial parameters which characterize the initial conditions, this lends credence to the assumption that the phase of active cluster formation (associated with warm dust and turbulence) is distinct from the phase of cold dust accumulation.  It also corroborates the notion that the cluster-forming clump is distinct from its stream of accreting matter. 

One further dimensionless parameter compares the rate of mass inflow to the rate of collapse for a virialized object: 
\begin{equation}\label{eq:xic}
\xic = {G \Mdotin \over \sigmac^3}. 
\end{equation}
This gives us some measure of the importance of inflow in the dynamical state of the clump.   The evolution of $\xic$ must depend on what controls the velocity dispersion $\sigmac$. If {\em inflow} drives turbulence, as pictured by \citet{2010A&A...520A..17K}, one might expect $\xic$ to approach a characteristic value determined by the properties of the inflow -- its density, velocity, and magnetic field structure, as well as its rate parameter ($\etaM$); {\bff see \S~\ref{SS:Accretion-energetics-stability}}.  

 Alternatively, if {\em feedback} is significant, then $\sigmac$ {\bff will reflect the driving of turbulence by stars as well as accretion, causing it to take a higher value for a given inflow rate, and causing $\xic$ to take a lower, possibly changing value.  (We shall see that protostellar outflows cause only gradual changes in $\xic$.)  Mass ejection does not change this conclusion, because we expect it to have very little effect on the infall rate $\Mdotin$.   } Taking $\xic$ to be constant or slowly varying is therefore reasonable, at least until other feedback effects become important.   Without further information, we expect $\xic$ to be of order unity in steady accretion in the absence of stellar feedback, and to be reduced below its ordinary value when feedback stirs turbulence.     

If we {\bff use the definition of $\xic$ to} evaluate the mass inflow rate using the simple filamentary infall model of \S~\ref{SS:Filaments}, we find that the clump is only moderately more turbulent than the filaments feeding it: 
\begin{equation} \label{eq:sigmac_from_sigmafil} 
\sigma_c = {1.37 (\Lambdafil \Nfil)^{1/2} \over \xic^{1/3} }\csfil. 
\end{equation} 

The dimensionless ratios defined above are enough to characterize aspects of its evolution, such as the relation between several of its internal time scales: 
\begin{equation}\label{eq:times} 
     {\sigmac\over \Rc}  \left(
       \begin{array}{c} \tffc \\ t \\ \Mc/ \Mdotstar
       \end{array}  \right)
       \\  
           = 
        \left(
       \begin{array}{c}  0.50\,\sqrt{\alphavir}     \\  { 5\etaM\over \epsinbar\alphavir \xic} \\ {0.50\,\sqrt{\alphavir}\over f_g \SFRff} \end{array} \right) .
\end{equation} 
Here we have introduced $\epsinbar =\Mc/M_{\rm in}$.   There is a reasonably clear separation of time scales: for fiducial values ($\alphavir=1.6$, $\xic =  \etaM=1$, $f_g =\epsin= \epsinbar = 0.8$, $\SFRff=0.03$) we see that $(\tffc::\Rc/\sigmac::t::\Mc/\Mdotstar) = (1::1.6::6.2::42)$.  For these parameters, the clump is eternally several crossing or free-fall times old. 

The comparison of time scales can also be interpreted in terms of the physical extent of the reservoir which feeds the clump.  If we are correct that the inflow of matter corresponds to a collapse from the initial state, and takes about one initial free-fall time at each radius, then $t\simeq\tff(r_0)$ where $r_0$ is the initial radius of this matter at $t=0$.  Given the value of $t/\tffc$ from equation (\ref{eq:times}), this implies a relation between the clump radius and the initial radius $r_0(t)$ of matter reaching it: 
\begin{equation}\label{eq:InitialRadius}
{r_0(t)\over \Rc(t)} \simeq 4.7{ \etaM^{2/3}\over \alphavir \epsinbar \xic^{2/3}}, 
\end{equation} 
which is 3.6 for the fiducial values listed above.  The clump's feeding zone is only a few times its own radius at any time.\footnote{\bff   This calculation neglects the difference in time to fall to $\Rc$ rather than $r=0$, and also the possibility that effects which blow out matter also affect the duration of infall.} 

Under certain conditions even fewer parameters are required, because the gas mass fraction $f_g$ can be expressed in terms of the others.  To see this, first compare the rates {\bff of star formation and mass accretion}:\footnote{We  use decimal coefficients for convenience; here $10.1 = 10^{3/2}/\pi$.} 
\begin{equation} \label{eq:MdotstarToMdotc}  
{\Mdotstar\over \Mdotc} = {10.1\, \SFRff\over  \alphavir^{3/2}}  \, {f_g \over  \epsin \xic}. 
\end{equation}  
But if we define  $q_*=d\ln\Mstar/d\ln \Mc$, then $\Mdotstar/\Mdotc = q_* \Mstar/\Mc =(1-f_g) q_*$; therefore \begin{equation}\label{eq:fg_solved}
f_g = {q_* \alphavir^{3/2} \epsin \xic \over 10.1\, \SFRff + q_* \alphavir^{3/2} \epsin \xic}.
\end{equation} 
If the dimensionless parameters we have listed are truly constant, so that the clump's growth is self-similar, then  $q_* = 1$ because stars form in lock step with the clump.  This, along with the other fiducial parameters, implies $f_g=0.84$, i.e. a stellar mass fraction of 16\%.   A value of this order is to be expected, given that star formation is assumed to be slow and the clump is permanently only a few free-fall times old. 

We pause to review the evolution with mass of clump properties obtained by assuming the dimensionless ratios like $\alphavir$ and $\xic$ are constant, or slowly varying, while $\Min\propto t^\etaM\propto t^{6/\krho-2}$: 
\begin{equation}\label{eq:scaling} 
       \left(
       \begin{array}{c} \Rc \\ \sigmac \\ \Sigmac
       \end{array}  \right) 
                \propto 
        \left(
       \begin{array}{c} \Mc^{2+\etaM\over3\etaM}  \\  \Mc^{\etaM-1\over3\etaM} \\ \Mc^{-{4-\etaM\over3\etaM}} \end{array} \right) 
          \propto 
        \left(
       \begin{array}{c} \Mc^{1\over 3-\krho}  \\  \Mc^{2-\krho\over 6-2\krho} \\ \Mc^{-{\krho-1\over 3-\krho}} \end{array} \right) 
\end{equation} 
where by $\Sigmac$ we mean the clump's mean column density $\Mc/(\pi \Rc^2)$.  If the initial reservoir has a constant column density ($\krho=1$) then $\Sigmac$ will be constant while $\sigmac$ increases ($\sigmac\propto \Mc^{1/4}$), whereas if the reservoir is like our filamentary infall model or like a singular isothermal sphere ($\krho=2$), then $\sigmac$ is constant while the column decreases ($\Sigmac\propto \Mc^{-1}$).  We expect $1\leq\krho\leq2$ to bracket the plausible range of values for the main accretion phase, and argued for the upper end of this range in \S~\ref{S:Accumulation}.   A strong drop in the mass accretion rate would correspond to  $\etaM\ll 1$.

\subsection{Protostellar population} \label{SS:protostellar-population}

Our parameterization of the growing clump allows us to estimate the  population of protostars within it.  
If we suppose that individual stars acquire their masses at an average rate $\mdotstar = \epsstar \xistar \cs^3/G$ (as a result of infall at a rate $\xistar \cs^3/G$ of which only $\epsstar$ lands on the star) then the average number of accreting protostars at any time is 
\begin{equation}\label{eq:N_protostars}
 {\Mdotstar\over\mdotstar} = {10.1\, \SFRff \over \alphavir^{3/2}} {f_g\over \epsstar \xistar} \left(\sigmac/\cs\right)^{3}. 
\end{equation} 
Adopting $\epsstar\xistar =0.5$ and the fiducial values of the other parameters, this becomes $\sigmac^3/(1.61\cs)^3$.   This result is quite sensitive to the Mach number of clump turbulence, but implies tens of accreting protostars in regions like NGC~1333.  

\subsection{Accretion-driven turbulence: energetics and stability} \label{SS:Accretion-energetics-stability}

The expectations laid out above rest on an underlying assumption: that molecular cloud accretion can sustain within it the turbulent, self-gravitating region we call the cluster-forming clump -- and further, that the clump's velocity dispersion $\sigmac$ is regulated by the dynamics of inflow, around some characteristic value.   Is this realistic?   

An energetic argument raises doubts.  If the accretion flow arrives with close to zero specific energy, as we argued in \S \ref{S:Accumulation}, then its inflow velocity will be close to the clump escape velocity $\vescc = (2G\Mc/\Rc)^{1/2}$, and the turbulent energy should be created at a rate comparable to the inflow of kinetic energy, $\Mdotin \vescc^2$.  If the virial parameter $\alphavir$ is constant then $\sigmac\propto \vescc$ and so the driving rate is of order $ \Mdotin \sigmac^2$.  However the rate of turbulent dissipation is of order $ \sigmac^5/G$ (assuming the clump is supported by supersonic turbulence).   The two rates balance at a specific value for which $\sigmac^3\simeq G \Mdotin$ (i.e. $\xic\simeq1$).  However, the balance is unstable: if the clump is smaller at a given time, then (assuming $\alphavir$ is the same) its $\sigmac$ is higher, and dissipation outpaces driving.  Conversely if the clump is expanded, driving outpaces dissipation, and in fact the study of accretion molecular clouds by  \citet{2011ApJ...738..101G} shows evidence of unstable behavior (their figure 2).    Although the outcome will be complicated by the turbulent nature of the flow and by the renewal of matter during accretion, this energetic instability invites a closer look.  Two points suggest that it does not invalidate the notion of a characteristic value of $\xic$. 

First, we have found that the feeding radius $r_0$ is only a few times larger than the clump radius, even when our parameters take their fiducial values.  Accretion shocks and turbulent dissipation involve a loss of energy, which requires an increase in the binding energy, so the clump clearly cannot grow to be comparable in size to its parent region.   This places an upper limit on excursions of  $\Rc$ and $\xic$ relative to the state $\xic=1$, and we view this as a consequence of the non-zero energy of the initial state. 

Second, a lower limit on $\Rc$ and $\xic$ comes from non-zero angular momentum in the initial state, which is a feature of any realistic scenario for mass accumulation.    If turbulent motions pervade the initial conditions with a virial parameter $\alphares$, then a parcel from radius $r_0$ has a characteristic angular momentum $|{\mathbf j}|\simeq (\alphares G \Mc r_0/5)^{1/2}) $  and hence, if $\mathbf j$ is conserved, would orbit at a radius $|{\mathbf j}|^2/(G \Mc) \sim (\alphares/5) r_0$.   While some of this angular momentum exists in random motions which can can cancel during infall, the cancellation cannot completely erase it.  \citet{2006MNRAS.373.1563K} analyze this cancellation in order to predict disk radii during massive star formation, and find  that it reduces $|{\mathbf j}|$ on each shell by at most a factor of two (their Appendix A and Table A1), leading to a minimum circularization radius $\sim (\alphares/25) r_0$.    While there is no evidence of overall rotational support within star clusters, we note that this limiting radius is proportional to $r_0$ and would therefore imply a constant value of $\xic$. 

These observations suggest that our assumption of a steadily growing clump with a characteristic value of $\xic$ is realistic. The process should be examined in greater detail, and for this reason we have begun a set of numerical experiments (Hansen et al.~2015, in prep.).  Note, also, that the models of \citet{2011ApJ...738..101G} show no evidence of the energetic instability in clouds with turbulent driving due to star formation.  We turn to this topic below.

\section{Feedback from protostellar outflows} \label{S:Feedback}

Protostellar outflows have long been recognized as a ubiquitous signpost of star formation \citep{1987ApJ...321..370H,1991MNRAS.252..442P,1986ApJ...301..398M,1996A&A...311..858B}.  As a source of outward momentum, they can eject matter from the sites of individual star formation \citep{1988ApJ...324..907M,1995ApJ...450..183N,1996ApJ...470.1001M,1998Natur.392..685V,1998ApJ...495..871L}
or the larger clumps in which they are embedded \citep{1984ApJ...277..634L,1986ApJ...306L..29L,1986ApJ...303L..11G,1994ApJ...423..310B}.  
They have also been implicated in the energization of turbulence on clump scales \citep{2005ApJ...632..941Q,2010MNRAS.409.1412G,2010ApJ...722..971C,2011ApJ...726...46N,2011ApJ...737...56N,2012MNRAS.420...10M,2013ApJ...774...22P}
although their influence does not extend to molecular cloud scales \citep[e.g.][]{2010ApJ...715.1170A}
as originally proposed \citep{1980ApJ...238..158N,1989ApJ...345..782M}.   The numerical simulations by several groups  \citep{2006ApJ...640L.187L,2006ApJ...646.1059C,2006ApJ...653..416C,2007ApJ...662..395N,2007Ap&SS.307...35F,2008ApJ...687..354N,2009ApJ...692..816C,2009ApJ...695.1376C,2010ApJ...709...27W,2011ApJ...740..107C,2012ApJ...747...22H,2013ApJ...766...97M} are broadly consistent with these observational findings, but see \citet{2007ApJ...668.1028B} for an opposing view and \citet{2014MNRAS.439.3420M} for important qualifications.    

The importance of protostellar outflow feedback in the evolution of a clump is determined largely by the mean protostellar wind momentum per unit mass, $\vchar$.  The net force due to all the winds within a clump is $\Mdotstar\vchar$, and, ignoring any dependence of $\vchar$ on stellar mass and environment, the momentum from a protostar of mass $m_*$ is $ m_*\vchar$.  
The actual value of $\vchar$ is quite uncertain; observational estimates are usually below $30\,\kps$, but \citet{2013AJ....145...94D} stress that momentum is often underestimated. 

\subsection{Outflow-driven turbulence}

A simple comparison of forces shows the potential importance of outflows in stirring turbulence: the characteristic turbulent acceleration for a clump with velocity scale $\sigmac$ is $\sigmac^2/\Rc$, corresponding to a characteristic force $M_g \sigmac^2/\Rc$.  Comparing the outflow force to this,  
\begin{eqnarray} \label{eq:outflows_versus_clumpaccel} 
{\Mdotstar \vchar \over M_g\sigmac^2/ \Rc} &=& {2.0\, \SFRff \,\vchar \over \alphavir^{1/2}\, \sigmac}  \nonumber\\
&=& {1.4\, \kps \over \sigmac} \left(1.6\over \alphavir\right)^{1/2} {\SFRff\over 0.03} {\vchar\over\, 30\kps}. 
\end{eqnarray} 
This suggests that, for our choice of fiducial parameters, outflows can be a strong influence in clumps with $\sigmac\lesssim 1.4\,\kps$; in more turbulent regions they are weak.  To put it another way, so long as the outflow force couples to turbulent motions on the clump scale, we expect outflows to sustain turbulence of order $1.7 \, \kps$ (for fiducial parameters) on their own.   Since this is equivalent to the level of accretion-driven turbulence when $\Mdotin = 700 \,\xic \,M_\odot/$Myr, both accretion and outflows feedback should be important in supporting regions that gain up to about a thousand solar masses per  million years.   We estimate that about a quarter of Galactic star formation occurs in such clusters (using the cluster birthrate model of \citealt{1997ApJ...476..144M} and assuming a common formation time of 1\,Myr). 

Equation (\ref{eq:outflows_versus_clumpaccel}) shows that, in the context of constant accretion, a primary influence of protostellar outflows is to modify the clump velocity dispersion and therefore to reduce the parameter $\xic$ relative to what it would be in the absence of feedback.  The virial parameter $\alphavir$ may also be shifted, although more subtly.   This observation corroborates our assertion in \S~\ref{S:Params} that dimensionless parameters like $\alphavir$ and $\xic$ are likely to be roughly constant during a period of accretion-fed growth. 

There are several features of protostellar outflows which should be accounted for when considering their affect on clump turbulence.

First,  outflows are discrete events with a characteristic momentum $\mbarstar \vchar$ (where $\mbarstar$ is the mean stellar mass) and a wide range of individual intensities.  This leads to a characteristic distance $\ell$ on which outflows' force is applied, and suppresses their effect on clump turbulence (relative to a simple estimate based on force balance, e.g. assuming expression  (\ref{eq:outflows_versus_clumpaccel}) to be unity) if $\Rc>\ell$.    This also introduces $\Mc/\mbarstar$ as a parameter which can affect the strength of feedback. 

Second, outflows are highly collimated: this mitigates the first effect by extending the reach of their momentum injection relative to a model in which they are spherical.   

Third, collimation also implies that some of the outflow momentum may be lost from the clump as outflows drive flows that escape the clump entirely. 

Fourth, any additional source of turbulence, such as the stirring by an accretion flow, will affect the scale on which outflows deliver their momentum, as this happens when outflow-driven motions decelerate to the local turbulent speed or wave speed.  

\begin{figure}
\includegraphics[scale=0.44]{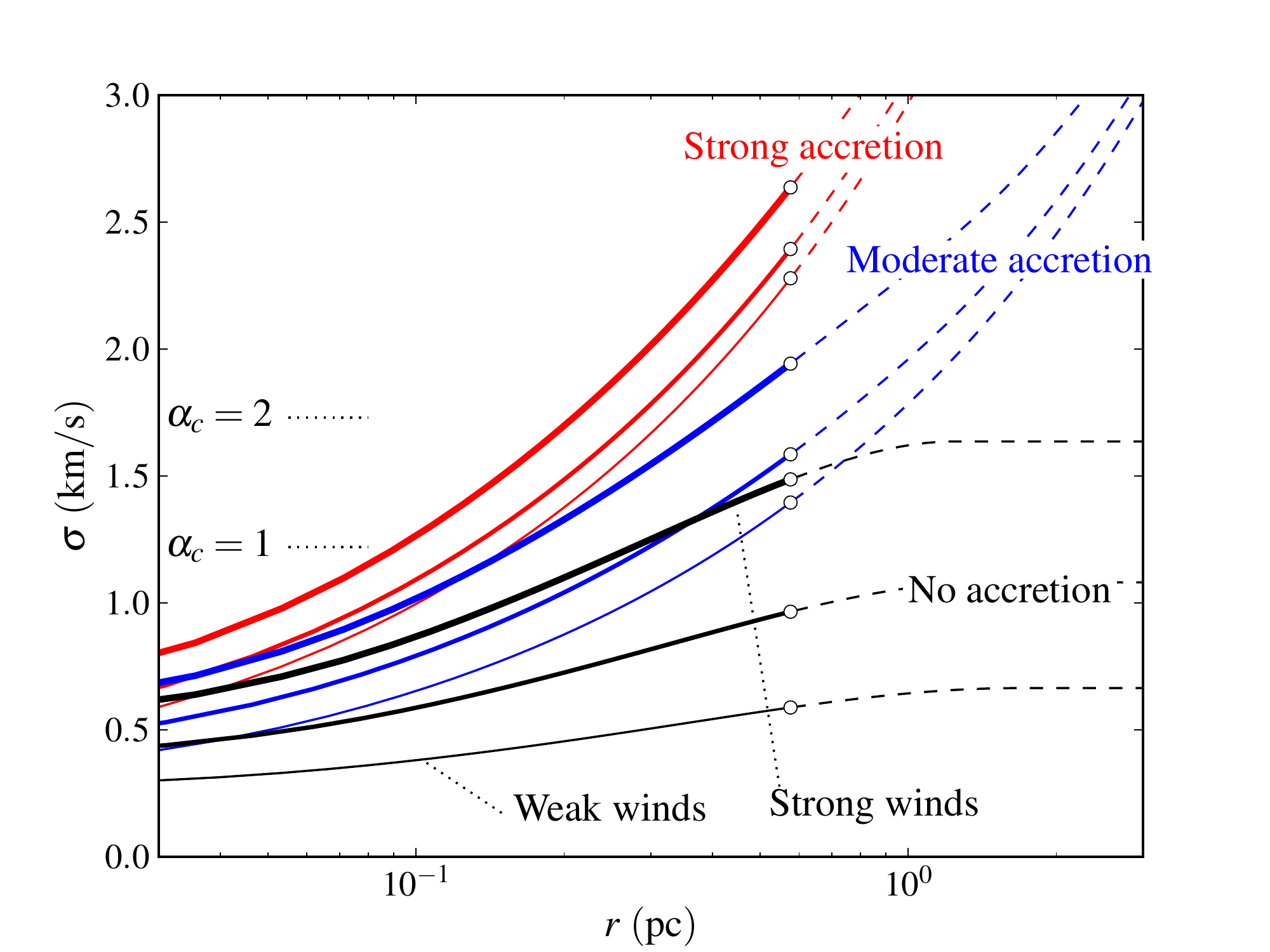}
\caption{Combined role of outflows and accretion in the M07 model.  A clump of 1000~$M_\odot$ with mean column density 0.2\,g\,cm$^{-2}$ is stirred by weak outflows ($\vchar = 5\,\kps$, thin lines), moderate outflows ($\vchar=15\,\kps$), or strong outflows ($\vchar = 40\,\kps$, thick lines), as well as no accretion ($\Mdotin=0$), moderate accretion ($\Mdotin=1000\,M_\odot$\,Myr$^{-1}$), or strong  accretion ($\Mdotin=3000\,M_\odot$\,Myr$^{-1}$).  Circles represent the clump radius, where the velocity dispersion $\sigmac$ and virial parameter $\alphavir$ are set.  Other parameters: $\SFRff = 0.034$; $\phiacc=0.75$; $\Lambda=1$; $c_s=0.19\,\kps$.  We adopt the Kroupa IMF and \citet{1999ApJ...526L.109M} collimation model, as in M07.} 
\label{Fig:ClumpExamples}
\end{figure}

\begin{figure}
\includegraphics[scale=0.44]{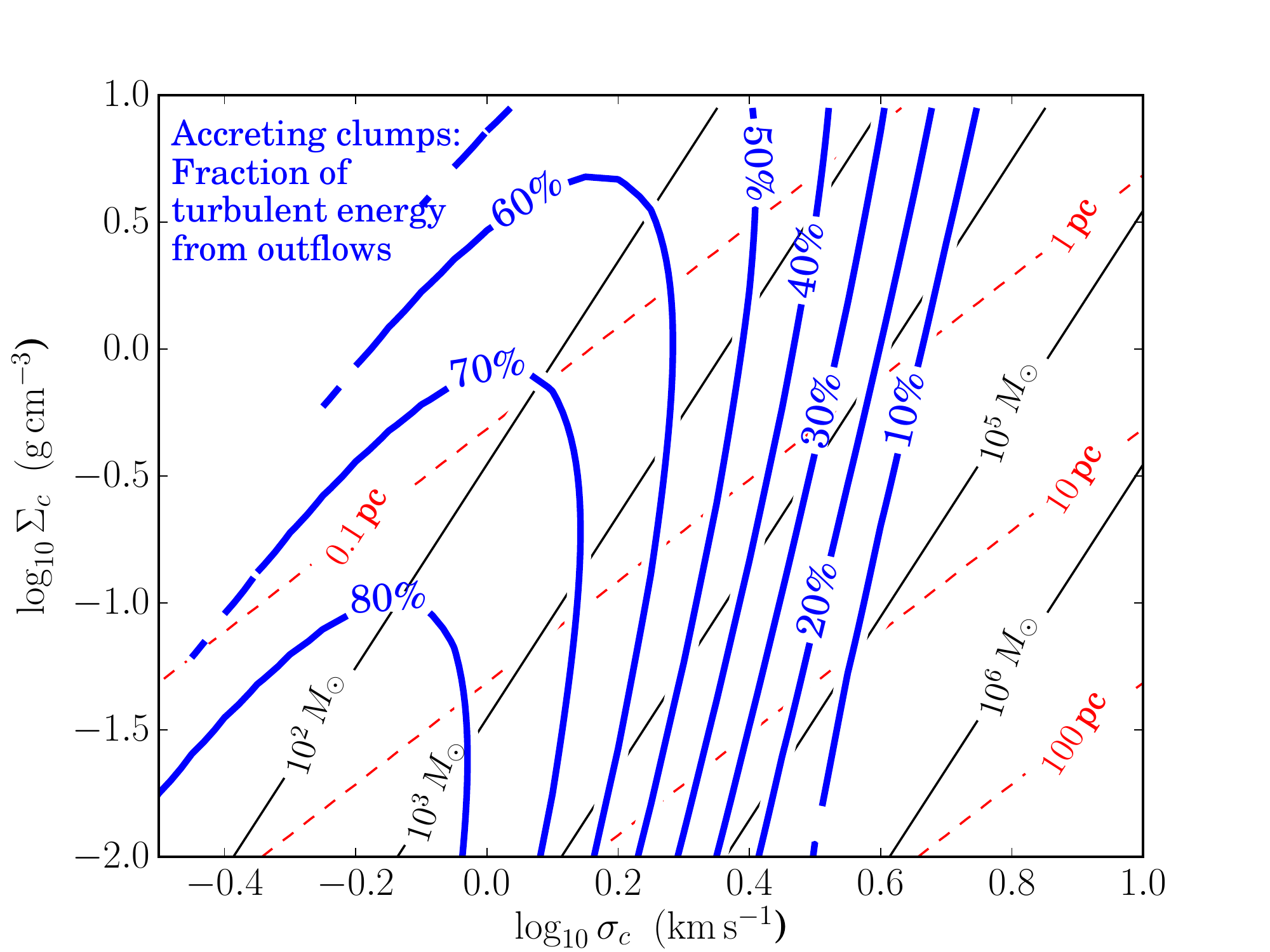}
\caption{Fraction of clump turbulence driven by outflows rather than accretion, $(1-\aext \Rc/\sigmac^2)$, in the M07 model of steady-state driving and decay.  This quantity is related to our parameter $\xic$ by equation (\ref{eq:xic-versus-phiacc}).  For each combination of $\sigmac$ and $\Sigmac$, we find the value of accretion driving ($\aext$) sufficient to maintain turbulence with virial parameter $\alphavir=1.6$ according to M07 equations (20), (21), and (22).  The M07 coupling parameter $\Lambda$ is set to 0.8, and other parameters take their fiducial values.  For $\Mc<100\,M_\odot$ fewer than fifteen outflows form per $\tffc$, so stochastic effects are strong.     } 
\label{Fig:AccretionOutflowPhasePlot}
\end{figure}

A simple model for outflow-driven turbulence that reflects all four effects was presented by \citeauthor{2007ApJ...659.1394M} (\citeyear{2007ApJ...659.1394M}, hereafter M07), and we shall use his equation (24) to estimate outflows' contribution to $\sigmac$.    We adopt the same stellar initial mass function and wind force structure function as M07 (from \citealt{2001MNRAS.322..231K} and \citealt{1999ApJ...526L.109M}, respectively) when evaluating his function ${\cal S}({\hat{ \cal I}})$ in M07's equation (30), and we also adopt his value of 0.8 for the outflow coupling efficiency (M07's parameter $\Lambda$).     The combined effect of accretion and outflows is depicted in figure \ref{Fig:ClumpExamples}, and in figure \ref{Fig:AccretionOutflowPhasePlot} we estimate the fraction of turbulent energy due to outflows, for accreting clumps with no other form of feedback.

The M07 theory allows us to account for the driving of turbulence by accretion, which influences how outflows interact with the gas.  For this we assume that accretion drives turbulence with an acceleration \[ \aext = \phiacc {\Mdotin \vescc\over f_g\Mc},\] where 
 $\phiacc\leq 1$ describes the efficiency with which accretion drives turbulence. (Our $\phiacc$ is similar to the parameter $\varphi$ introduced by  \citealt{2011ApJ...738..101G}.)   This definition, along with the definitions of $\alphavir$ and $\xic$, implies 
 \begin{equation} \label{eq:xic-versus-phiacc} 
\xic =  \left(5\over 2\alphavir\right)^{1/2} {f_g\over \phiacc} {\aext \Rc \over  \sigmac^2}, 
 \end{equation}  
and we use this relation to evaluate the relative influence of accretion and protostellar outflows.   When feedback is insignificant the M07 theory implies $\sigmac^2 = \cs^2 + \aext \Rc$, so that $\xic$ takes the unique value $(2.5/\alphavir)^{1/2}f_g \phiacc^{-1}$, assuming $\cs^2$ can  be ignored.   Dynamically significant outflows stir higher-velocity turbulence, leading to  an appreciable drop in $\xic$.   Equation (\ref{eq:xic-versus-phiacc}) can be evaluated if the fraction of turbulence supplied by outflows, $(1-\aext\Rc/\sigmac^2)$, is known; we plot this quantity in Figure \ref{Fig:AccretionOutflowPhasePlot} using the M07 theory.    The result is approximately consistent with our expectation, from equation (\ref{eq:outflows_versus_clumpaccel}), that outflows are strong when $\sigmac\lesssim 1.4$\,km\,s$^{-1}$, but it is modified somewhat by their collimation and escape.  For instance, the loss of momentum in escaping outflows is very significant when $\Mc\lesssim 100\,M_\odot$. 

\subsection{Outflow driving: energetics and stability} \label{SS:outflows-energetics-stability} 

If protostellar outflows are capable of sustaining a virial level of turbulence within a clump, is this balance energetically stable?  A simple argument, like the one considered in \S~\ref{SS:Accretion-energetics-stability}, shows that the answer is no.  The rate of turbulent energy injection by protostellar outflows is $\sim \Mdotstar \vchar \sigmac \sim \SFRff \sigmac^4\vchar/ G$, while the rate of turbulence dissipation is $\sim \sigmac^5/G$.  Unless there is an effect which causes $\SFRff$ to vary rapidly with $\sigmac$, a balance between these is unstable to variations of $\sigmac$ away from the equilibrium state.   Star formation is observed to be suppressed for low values of the visual extinction (i.e. low $\Sigmac$), but this threshold is well below the columns of interest in star cluster formation.   

We infer, therefore, that {\em a clump supported entirely by outflow-driven turbulence is energetically unstable} and should oscillate  around its equilibrium state, or explode or collapse altogether.   The discreteness of individual outflows makes this behavior stochastic, especially in small clumps, as the number of stars formed per free-fall time is $\sim 12(\Mc/100\,M_\odot)$.   (Note that  \citealt{1999PhDT........11M} argues that this instability leads to over-stable oscillations.)

Similar conclusions hold for feedback effects like H\,II regions within giant molecular clouds, which also form at a rate proportional to the star formation rate, inject a reasonably constant momentum per unit mass, and involve discrete events.   Comparing the results of \citet{2006ApJ...653..361K}, who do not include accretion, and  \citet{2011ApJ...738..101G}, who do, we infer that the instability is damped by accretion.  

{\bff The above argument} rests on the negative specific heat of spherical self-gravitating systems: if a loss of energy led $\sigmac$ to decline, we would have inferred stability rather than instability.   Filamentary and planar systems (such as the disks of spiral galaxies) do not have negative specific heats, so stellar feedback leads to stable equilibria in these systems.  We have simplified a complex system into just a few degrees of freedom, so we cannot draw firm conclusions on the outcome of the instability without conducting numerical experiments.

\subsection{Outflow mass ejection} \label{SS:outflowejection} 

\begin{figure}
\includegraphics[scale=0.44]{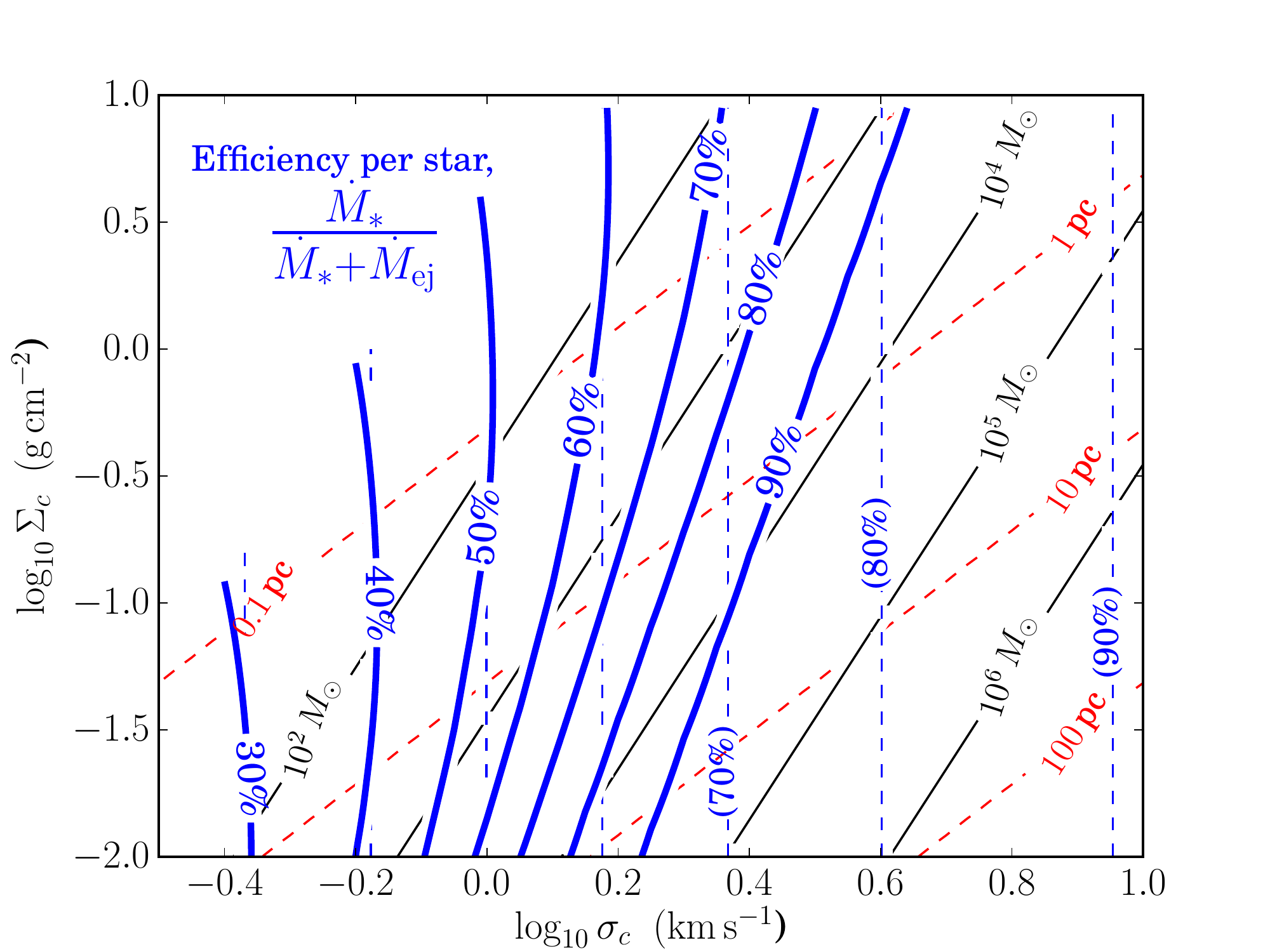}
\caption{Efficiency of star formation, $\Mdotstar/(\Mdotstar+\Mdotej)$, considering only mass ejection due to outflows.  Solid blue lines are an evaluation of equation (\ref{eq:MdotejOnMdotstarFromM07}), which accounts for the effect of outflow confinement in large clouds.  Thin blue lines (with parenthesized values) represent equation (\ref{eq:MejPerStar}), an approximation from MM00, which does not.    {\bff Efficiency increases with $\sigmac$ because each star ejects less mass in a region of higher escape velocity. }  These curves assume a \citet{2001MNRAS.322..231K} initial mass function and a \citet{1999ApJ...526L.109M} outflow structure with core angle $\theta_0=10^{-2}$. }
\label{Fig:EfficiencyPhasePlot}
\end{figure}

Strong collimation allows outflows to breach the clump and eject matter.  As \citeauthor{2000ApJ...545..364M} (\citeyear{2000ApJ...545..364M}, hereafter MM00) explain, an absolute upper limit for the mass ejection rate $\Mdotej$ produced by outflows is obtained imagining that the outflow momentum couples perfectly to motions just fast enough to escape the clump.  Since each star emits momentum $m_* \vchar$, the upper limit of mass ejected by this one star is  $M_{\rm ej,*} = m_* \vchar/\vescc$.   MM00 show that, in the \citet{1999ApJ...526L.109M} outflow model, the actual ejected mass is lower by a constant factor,  for each outflow  strong enough to break free of the clump, but too weak to entirely disrupt it.  Taking this factor into account, 
\begin{equation}\label{eq:MejPerStar}
{M_{\rm ej, * }\over m_*} =  {1\over 2c_g \ln(2/\theta_0) }{\vchar \over \vescc} \simeq {\vchar\over 12.0\,\vescc}
\end{equation} 
where $\theta_0 \simeq 10^{-2}$ defines a core angle for the outflow, and  $c_g \simeq 1.13$ corrects for deceleration as an outflow crosses the clump.   {\bff The ejected mass per star is inversely proportional to $\vescc$, implying that the star formation efficiency increases with $\sigmac$ as seen in Figure~\ref{Fig:EfficiencyPhasePlot}. }

Since this ratio is independent of stellar mass, for outflows in this intermediate range, it provides an estimate of the ratio  $\Mdotstar/\Mdotej$ between star formation and mass ejection.    An improved estimate must account for other effects, like the confinement of weak outflows within giant clouds, which reduce $\Mdotej$.  MM00 evaluate these effects numerically (their figure 2), but we note that they can also be represented using the functions defined by M07: 
\begin{equation}\label{eq:MdotejOnMdotstarFromM07}
{\Mdotej\over \Mdotstar} = {f_g \Mc \over \mbarstar} \left[1 -{ \hat {\cal S}({\cal I}_{\rm esc}) \over \hat {\cal S}_{\rm tot}}\right]. 
\end{equation} 
Here ${\cal I}_{\rm esc} = c_g f_g \Mc \vescc$ is the impulse required to unbind all the gas from the clump (which is also the isotropic-equivalent impulse required to unbind gas in one direction).   The M07 function $\hat {\cal S}(\hat {\cal I})/{\cal S}_{\rm tot}$ is a normalized cumulative distribution of outflow strengths -- or more precisely, the strengths of all the individual angular sectors of all the outflows -- emitted in the creation of a stellar population (M07 equation 30).    In figure \ref{Fig:EfficiencyPhasePlot} we compare expression (\ref{eq:MdotejOnMdotstarFromM07})  to equation (\ref{eq:MejPerStar}), which provides an approximation for the same quantity. 

We now wish to revisit the conditions of an accreting clump for the case in which there is a relation between the rates of mass ejection and star formation, as there is when protostellar outflows are the cause.  Using the equality $\Mdotin = \Mdotc + \Mdotej$ with  equation (\ref{eq:MdotstarToMdotc}), we find 
\begin{equation}\label{eq:epsin_evaluated} 
\epsin = 1- Y  \left(\Mdotej\over \Mdotstar\right)f_g
\end{equation} 
where $Y=10.1\,\SFRff/(\alphavir^{3/2} \xic)$.  The clump experiences a net loss of mass ($\epsin<0$) if ejection is stronger than accretion, i.e. if  $\Mdotej/\Mdotstar >1/Y$, which is $6.7(1.6/\alphavir)^{3/2}$ for our fiducial parameters.   

\begin{figure}
\includegraphics[scale=0.44]{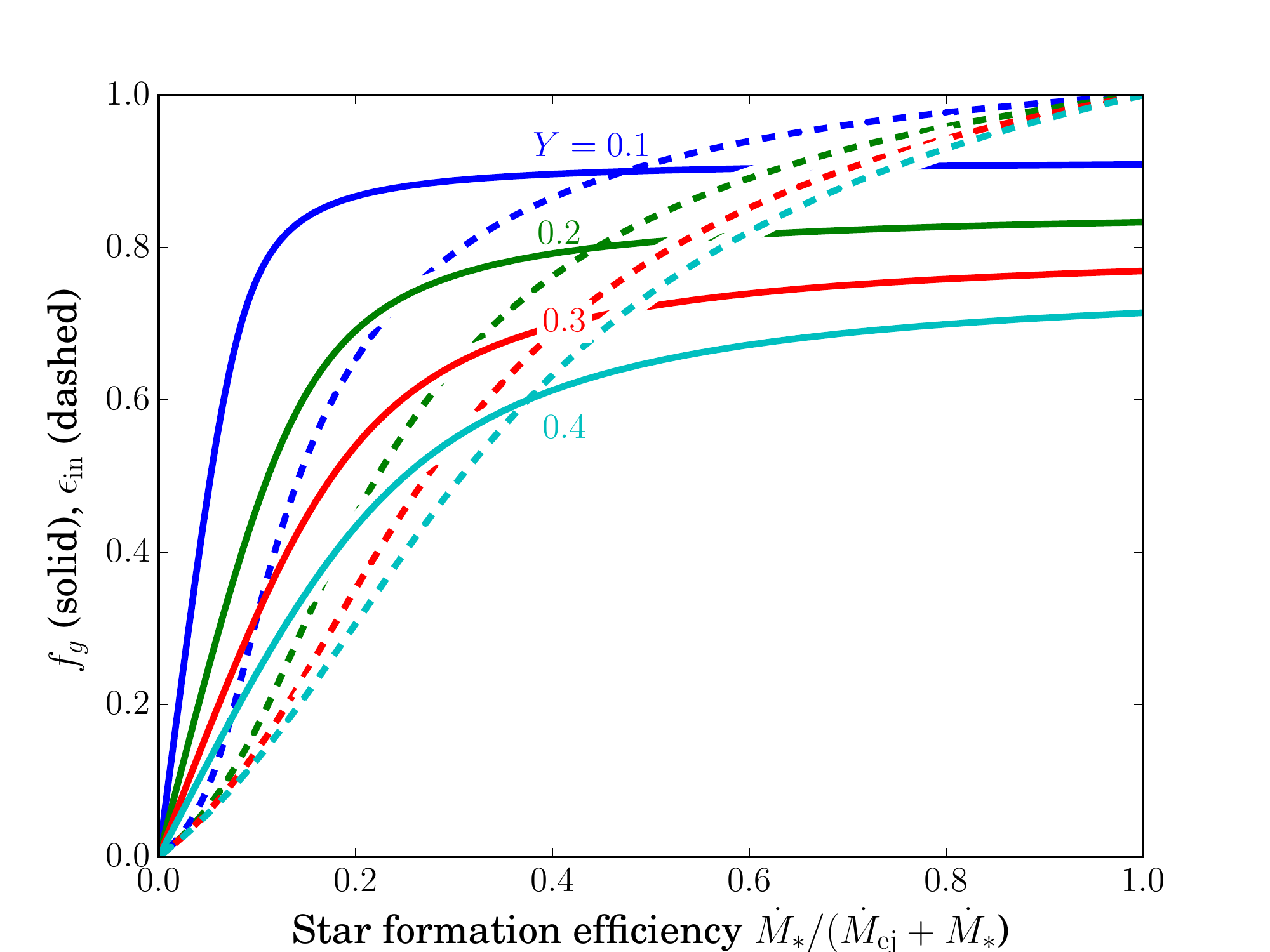}
\caption{Gas fraction $f_g$ (eq.~\ref{eq:fgFromMdotejPerStar}, solid lines) and accretion efficiency $\epsin$ (eq.~\ref{eq:epsin_evaluated}, dashed lines) as functions of the star formation efficiency and the parameter $Y=10.1\,\SFRff/(\alphavir^{3/2} \xic)$, for clumps undergoing strictly self-similar growth ($q_*=1$).}
\label{Fig:fg_and_epsin}
\end{figure}

For the case of a clump undergoing self-similar growth, we can go further by employing equation (\ref{eq:fg_solved}), in which we use the definition $q_* = d\ln \Mstar/d\ln \Mc\simeq 1$ to relate $f_g$ and $\epsin$.  This shows that the gas fraction $f_g$ can be determined as the physically relevant solution of the quadratic equation $(1-f_g)q_*[1-Y (\Mdotej/\Mdotstar) f_g ]= Y f_g$: 
\begin{equation}\label{eq:fgFromMdotejPerStar}
f_g = {W - \sqrt{W^2 - 4 (\Mdotej/\Mdotstar) q_*^2 Y} \over 2 (\Mdotej/\Mdotstar) q_* Y },
\end{equation} 
where $W = q_* + Y[1+(\Mdotej/\Mdotstar) q_*]$.     This result and the corresponding value of $\epsin$ are plotted in figure \ref{Fig:fg_and_epsin} for the case $q_*=1$.  

\begin{figure}
\includegraphics[scale=0.44]{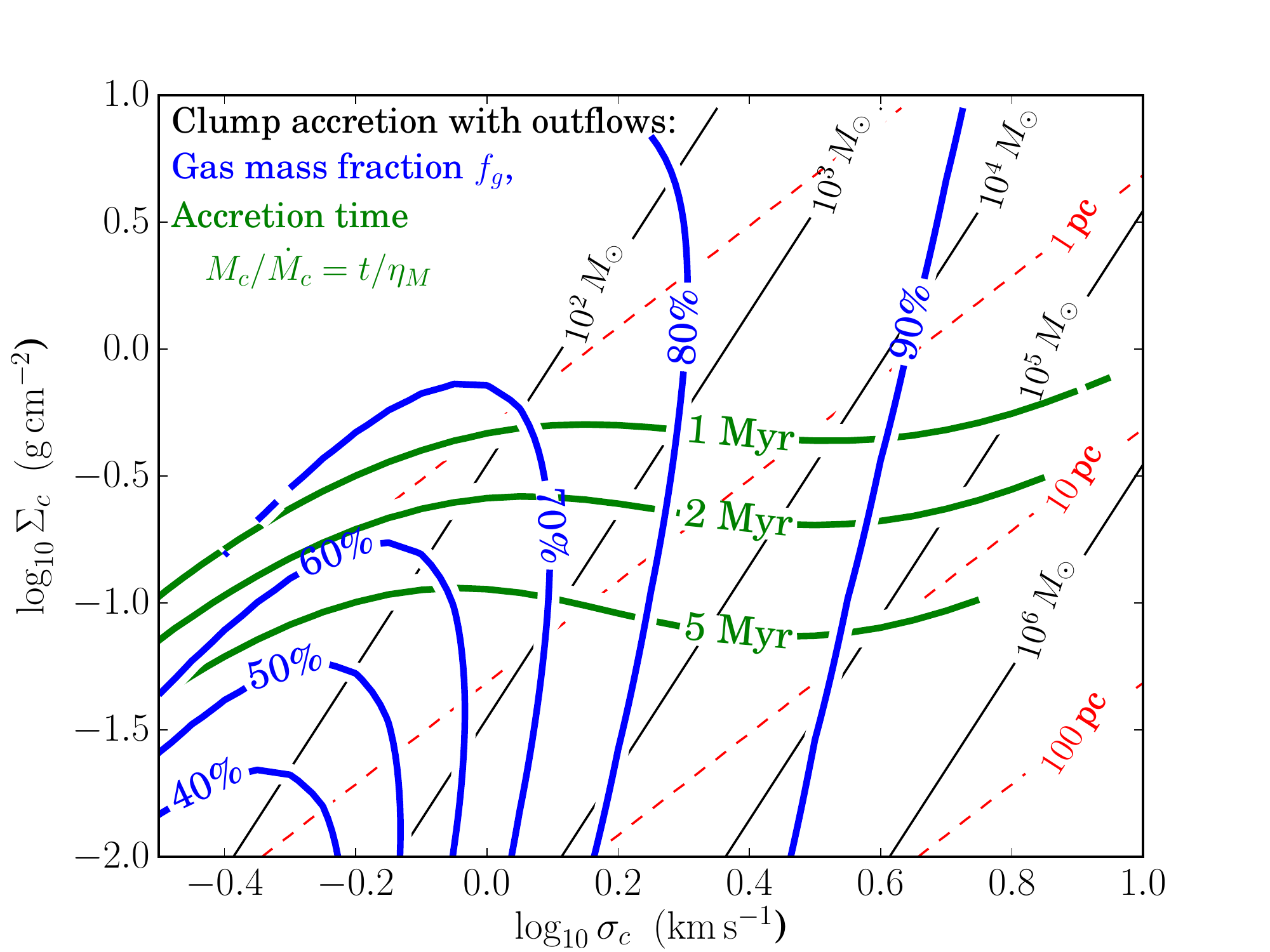}
\caption{Gas fraction $f_g$ and accretion time $\Mc/\Mdotc = t/\etaM$ for clumps accreting self-similarly under the influence of protostellar outflows (i.e., neglecting massive-star feedback).   Outflows affect these curves both by ejecting matter (reducing $\epsin$) and by injecting turbulent energy (increasing $\sigmac$ and reducing $\xic$). Other parameters ($\vchar$, $\alphavir$, $\phiacc$, $\SFRff$) are held
 fixed at fiducial values, and self-similar growth ($q_*=1$) is assumed in the calculation of $\epsin$ and $f_g$.   }
\label{Fig:fgPhasePlot}
\end{figure}

In Figure \ref{Fig:fgPhasePlot} we combine these results to make a prediction for the gas fraction $f_g$ and accretion time scale $\Mc/\Mdotc=t/\etaM$ for clumps undergoing self-similar accretion whilst being afflicted by outflows (and no other feedback).  Beginning with the determination of $\Mdotej/\Mdotstar$ (eq.~\ref{eq:MdotejOnMdotstarFromM07}) and $\aext\Rc/\sigmac^2$ from the M07 theory, we calculate $\xic$ via equation (\ref{eq:xic-versus-phiacc}) and then $\epsin$ and $f_g$ from equations (\ref{eq:epsin_evaluated}) and (\ref{eq:fgFromMdotejPerStar}).   The accretion rate is evaluated using $\Mdotc = \epsin \Mdotin = \epsin \xic \sigmac^3/G$, and used to compute the accretion time scale. 

One interesting feature of outflows' influence that is visible in this figure is that the clump age and accretion time are primarily a function of the clump column $\Sigmac$.   For $\Mc/\Mdotc \simeq 1$\,Myr the characteristic column is about 0.3\,g\,cm$^{-2}$.    

The path of an accreting clump through this diagram is determined by the accretion history: for self-similar steady accretion ($\krho=2$, $\etaM=1$) a clump moves downward at nearly constant $\sigmac$; for accelerating accretion ($\krho=1.5$, $\etaM = 2$) the motion is down and to the right, at roughly constant $\sigmac^2\Sigmac$; and for strongly accelerating accretion ($\krho=1$, $\etaM=4$) it is horizontally to the right, at approximately constant $\Sigmac$.   

\section{Massive stellar feedback} \label{S:Radpressure}

Protostellar outflows are relatively unimportant in  clusters with velocity dispersions $\sigmac>2\,\kps$.  Higher velocity  dispersions are the province of massive clusters containing B and O stars, which afflict cluster-forming gas with photon pressure, photo-ionization, stellar winds, and eventual supernovae.  The relative importance of these effects on dense cluster-forming gas has been discussed several times (e.g. \citealt{2009ApJ...703.1352K}, \citealt{2009ApJ...691..946M}, \citealt{2010ApJ...709..191M},  \citealt{2010ApJ...710L.142F}, and \citealt{2014MNRAS.442..694D}, among others).   Our goal is to review and reconsider these effects in the context of accretion onto the cluster-forming region.  In particular, we wish to delineate regions in the parameter space of Figure \ref{Fig:fgPhasePlot} where each effect is most important, and estimate the implications for gas removal.    

Our observational points of reference are the cluster-forming regions traced in dust continuum, high-density molecular tracers, and signposts of massive star formation such as water masers, methanol masers, or compact/ultra-compact H\,II regions \citep[e.g.,][]{1997ApJ...476..730P, 2003ApJS..149..375S, 2003A&A...410..597W,2004A&A...426...97F,2005ApJ...635L.173W,2011ApJ...741..110D,2013MNRAS.431.1752U}.  These regions span a wide range of gas mass ($10^2-10^4\,M_\odot$) with column densities $\sim 0.6$\,g\,cm$^{-2}\pm 0.38$\,dex (in the \citeauthor{2003ApJS..149..375S}\ sample); their infrared emission reflects characteristic dust temperatures $\sim 32-40$\,K and luminosity-to-mass ratios $\sim 70-130\,L_\odot/M_\odot$ \citep{2002ApJS..143..469M,2002ApJ...566..945B, 2004A&A...426...97F}.   They are about as likely to show evidence of inflow, in the line profiles of self-absorbed tracers, as are individual low-mass protostars, albeit at much higher flow rates \citep{2003ApJ...592L..79W,2015MNRAS.450.1926H}. 

Once sufficiently many stars are born, the complement of massive stars grows, and the strength of their feedback grows as well.   Massive-star feedback is initially stochastic, as it depends strongly on the mass of the most massive star in the population, but it becomes more predictable once the cluster samples the initial mass function (IMF) to the highest stellar masses.   We illustrate this in figures \ref{Fig:ClusterSampling} and \ref{Fig:SNsampling}, which describe the growth of the initial stellar luminosity-to-mass ratio $L_*/M_*$ (to its IMF-averaged value $\sim 10^3L_\odot/M_\odot$) and the probability of experiencing a core-collapse supernova (SN), respectively, during the growith of a cluster.   Photo-ionization depends on the production rate $S$ of H-ionizing photons, which grows to  $S\simeq L_*/(47\,{\rm eV})$ for clusters which sample the entire IMF. (For this estimate we use zero-age main sequence luminosities from \citealt{1996MNRAS.281..257T} and ionization rates from \citealt{1996ApJ...460..914V}.)   Once its stars have formed, a cluster's luminosity and ionizing output will be constant for the main-sequence lifetime of its most massive stars, which is most of the time prior to its first supernovae (a few Myr).    The detailed properties and evolution of a massive cluster vary somewhat depending on the metallicity, stellar multiplicity \citep{2012Sci...337..444S}, stellar rotation \citep{2013ApJ...764...21C}, line blanketing in stellar winds \citep{2005A&A...436.1049M}, and the form of the stellar initial mass function; however we do not attempt to capture these effects. 

We pause to show that, when they exist,  massive stars are much brighter than the accretion and contraction luminosity of the low-mass stars.  At stellar ages $t_*\sim $  Myr, low-mass stars are fully convective objects with surface temperatures $T_{\rm eff}\simeq 4120\,\hat m_*^{0.13}$\,K at stellar mass $\hat m_*\,M_\odot$ \citep{2009ApJ...702L..27B}.  The luminosity of such an object, averaged over its short life, is its binding energy divided by its age, which we compute \citep[e.g.,][]{1998ApJ...497..253U} to be $5.2\,\hat m_*^{0.51} ({\rm Myr}/t_*)^{2/3}\,L_\odot/M_\odot$.  Making the simplifying assumption that $m_*$ is chosen from the IMF, we find that the mean luminosity per unit mass of the young stellar objects is \[(1.8~{\rm to~} 2.8)\left< \left({\rm Myr}\over t_*\right)^{2/3}\right> {L_\odot\over M_\odot},\] where $\left<\,\right>$ indicates an average over stellar ages, and the range of prefactors reflects the outcome from different IMFs.   

The accretion and contraction luminosity of the clump are smaller still, and entirely negligible.
  
\subsection{Radiation force: direct and indirect} \label{SS:Frad} 
To begin, we compare the direct radiation force against gravity.  Because dust illuminated directly by un-extinguished starlight has temperatures $\gtrsim 150$\,K,\footnote{We arrive at this estimate by assuming a dust opacity to starlight of $\sim 10^{-21}$ cm$^2$ per H atom \citep{2011ApJ...732..100D}, Planck mean opacity from \citep{2003A&A...410..611S}, and the starlight flux evaluated at $r<\Rc/3$. }
 and because the clump is optically thick (optical depth $\sim 20 f_g \Sigmac/$(g\,cm$^{-2}$)) to the mid-infrared radiation emitted by these grains, we include this first stage of reprocessing in the `direct' radiation force, boosting it by the factor $1+\phiIR\simeq 2$. (Further reprocessing is non-local and best incorporated into the `indirect' force discussed below.)  For simplicity, assume all the gas is swept into a shell of radius $\Rc$.   The weight of the gas shell is $f_g (1-f_g/2) G \Mc^2/\Rc^2$, and the photon force (including  mid-IR reprocessing) is $L_*/c=(1+\phiIR)(1-f_g)(L/M)_* \Mc/c$, so the ratio of the direct radiation force to the weight of gas is 
\[ F_{\rm rad, dir}/F_{\rm grav} = {(1+\phiIR)\Sigmacrit  /\Sigmac},\]
  where 
\begin{eqnarray} \label{eq:Sigmacrit} 
\Sigmacrit &=&{1/f_g-1 \over 1-f_g/2} {(L/M)_*\over \pi G c} \nonumber \\
 &=& 0.31 \left[{1/f_g -1 \over 1-f_g/2} \right]{(L/M)_*\over 10^3\,L_\odot/M_\odot } {\rm g\,cm^{-2}}.
\end{eqnarray} 
The factor in brackets is unity for $f_g = 2-\sqrt{2} = 0.59$, and increases rapidly for smaller gas fractions.  

Next we consider the indirect force, which has been approximated in different ways. 
 \citet{2009ApJ...703.1352K} define  $\ftrapr= F_{\rm rad}/(L/c)$, and employ a leaky-shell model to estimate $\ftrapr \lesssim 1$. 
\citet{2010ApJ...709..191M}  consider instead a closed spherical shell of infrared optical depth $\tau_{\rm IR}=\int \bar \kappa(r) \rho(r) dr$ (if $\bar \kappa(r)$ is the flux-averaged opacity),  and so estimate  $\ftrapr = \tau_{\rm IR}$ and $F_{\rm rad, ind} = \tau_{\rm IR}L/c$.    Even more sophisticated approaches \citep[e.g.,][]{2005ApJ...631..792C} involve radiative transfer solutions in spherical symmetry.  
We opt for simplicity, on the basis that the local Eddington factor 
\begin{equation} \label{eq:Gammaedd}
\Gamma(r) = {\bar\kappa(r) L(r) \over 4\pi G M(r) c}= \frac{L(r)/M(r)}{10^3L_\odot/M_\odot} \frac{\bar \kappa}{13\,{\rm cm^2\,g^{-1}}},  
\end{equation} 
which measures the ratio of radiation force to gravity (for flux-averaged opacity $\bar \kappa$), is modest and bounded.    In an optically thick region $\bar \kappa$ is the Rosseland mean; Rosseland mean opacity in the models of \citet{2003A&A...410..611S} peaks at a maximum value of $3-10$\,cm$^2$\,g$^{-1}$, at temperatures $\sim 100$\,K (just cool enough for ices to persist).    Considering that $M(r)$ includes the mass of gas as well as stars,  that $(L/M)_*$ saturates at $\sim 10^3\,L_\odot/M_\odot$, and that clump dust often temperatures well below $100$\,K, typical values of $\Gamma$ are less than unity.  

Indeed it suffices, for our purposes, to assign a single average dust temperature $T_d$: the value for which 
\begin{equation}\label{eq:Td_appx} 
{\sigma_{r} T_d^4 \bar \kappa_{\rm Pl}(T_d) \over 1+3\bar \tau_{\rm Pl}(T_d) \bar \tau_{\rm R}(T_d)/8 } = {L\over 4 M_g} = {1-f_g\over4 f_g} \left(L\over M\right)_*, 
\end{equation} 
where $\sigma_r$ is the Stefan-Boltzmann constant and $\bar \tau_{\rm Pl} = \bar \kappa_{\rm Pl} \Sigmac$ and $\bar \tau_{\rm R} = \bar \kappa_{\rm R} \Sigmac$ are the mean Planck and Rosseland optical depths, respectively.   In the optically thin limit, this gives the dust temperature required to radiate the given luminosity per unit mass, $\sigma_r T_d^4 = L/(4 M_g \bar \kappa_{\rm Pl} )$, which correctly reproduces 30\,K dust for the median source in the \citeauthor{2004A&A...426...97F}\ sample.  In the optically thick limit, it approximates the radiation diffusion equation with $\sigma_r T_d^4  = (3/8) \bar\tau_{\rm R} L/(4\pi \Rc^2)$; the factor $3/8$ derives from a slab model.

To be specific, we adopt the model of composite-aggregate grains with normal silicates from \citet{2003A&A...410..611S}, for which $\bar \kappa_R \simeq 3.0 (T_d/100\,{\rm K})^{1.93}$\,cm$^2$\,g$^{-1}$ and  $\bar \kappa_{\rm Pl}\simeq 6.1(T/100\,{\rm K})^{1.58}$\,cm$^2$\,g$^{-1}$ for $30<T_d<100$\,K; for 100-700\,K, $\bar \kappa_{\rm R}$ and $\bar \kappa_{\rm Pl}$ are reasonably constant.  The temperatures in the optically thin and thick limits are therefore 
\[ T_{d,{\rm thin}} = 47 \left(L\over M_g\right)_3^{0.18}\,{\rm K}\] 
 and
  \[T_{d,{\rm thick}} = 32 \left(L\over M_g\right)_3^{0.48} \Sigma_{g,{\rm cgs}}\,{\rm K},\] 
 respectively, so long as $T_d<100$\,K, and the transition from thin to thick occurs for $\Sigma_{g,{\rm cgs}} = 1.46 (L/M_g)_3^{-0.31}$.  (Here we employ shorthand: subscript `3' on $L/M$ means units of $10^3 L_\odot/M_\odot $, and `cgs' means cgs units, e.g. g\,cm$^{-2}$.)   A good approximation to the solution of equation (\ref{eq:Td_appx}) is 
\begin{equation}
T_d \simeq (T_{d,{\rm thin}}^{3.1} + T_{d,{\rm thick}}^{3.1})^{1/3.1}. 
\end{equation} 

We then estimate $\Gamma$ using an estimate of $\bar\kappa$.   For the optically thick case $\bar \kappa=\bar \kappa_{\rm R}(T_d)$.  For an optically thin clump, the appropriate $\bar \kappa$ is the opacity of dust to the glow of other grains at temperature $T_d$:  $\bar \kappa = \bar\kappa_{\rm dd}(T_d)$, where 
\[\bar \kappa_{dd} = {\int_0^\infty B_\nu(T_d) \kappa_\nu^2\, d\nu \over\ \int_0^\infty B_\nu(T_d) \kappa_\nu\, d\nu}.\]    
In the adopted dust model, $\bar \kappa_{dd}\simeq 9.0 (T_d/100\,{\rm K})^{0.92}$\,cm$^2$\,g$^{-1}$ for $30<T_d<100$\,K.   Combining the thin and thick limits and using equation (\ref{eq:Gammaedd}), we estimate 
\begin{equation} \label{eq:GammaEdd_approx}
\Gamma \simeq \Gamma_{\rm thin} + \Gamma_{\rm thick}
\end{equation}
where 
\begin{equation} \label{eq:GammaThin} 
\Gamma_{\rm thin} = 0.34 f_g (f_g^{-1}-1)^{1.16} (L/M)_{*,3}^{1.16}
\end{equation} 
and 
\begin{equation} 
\Gamma_{\rm thick} = 0.026 f_g (f_g^{-1}-1)^{1.93}  (L/M)_{*,3}^{1.93} \Sigma_{c, {\rm cgs}}^{1.86}. 
\end{equation} 
These expressions are only valid for $T_d<100$\,K.  The maximum value of $\Gamma$ (corresponding to $T_d=100$\,K) is 0.23\,$(L/M_g)_3$, in the optically thick limit, and 0.69\,$(L/M_g)_3$ in the thin one.  Note that $\Gamma$ depends on the stellar IMF, through $(L/M)_{*,3}$,  and on the dust properties.  For  instance, \citeauthor{2003A&A...410..611S}'s `homogeneous aggregates' are several times more opaque, and can produce $\Gamma\simeq 1$ for reasonable stellar IMFs. 

Because $\Gamma$ is the negative ratio of the indirect radiation force to gravity, for a parcel of dusty gas, we can account for the indirect force by replacing Newton's constant $G$ with $G' = (1-\Gamma)G$ in any expression which involves the weight of the gas. For clarity, we add a prime to any quantity (e.g., $\alphavir'$, $\tffc'$) that we modify in this way. 

\subsection{Stellar winds} \label{SS:Winds} 

\citet{2010ApJ...709..191M} provide strong arguments that the pressure of shocked stellar winds can never build up enough to overwhelm the direct radiation pressure and photo-ionized gas pressure: leakage of hot gas, or radiative losses enhanced by thermal conduction, sap its strength.   Observations of H\,II regions corroborate this point \citep{2009ApJ...693.1696H, 2011ApJ...731...91L, 2011ApJ...738...34P, 2012ApJ...757..108Y}, although these generally address exposed regions on scales larger than our clumps.  

We will therefore  adopt the conservative estimate that stellar winds transmit no more force to the clump than is imparted to them at the stellar surfaces, 
\begin{equation} 
F_w = \phi_w {L\over c}
\end{equation} 
where $\phi_w\simeq 0.5$ \citep{2009ApJ...703.1352K}.   By this estimate, wind force is smaller than the force due to mid-infrared reprocessed starlight.

\subsection{Photo-ionization} \label{SS:Photoionization} 

The pressure of photo-ionized gas (H\,II) is a potentially important driver of clump motions and of mass loss, which can occur either through dynamical ejection of neutral gas or the evaporation of ionized gas.   In the following discussion we adopt an ionized gas temperature $T_i =10^4 T_{i,4}$\,K, a case-B recombination coefficient  $\alpha_B = 2.63\times10^{-13}T_{i,4}^{-0.8}\,$cm$^3$\,s$^{-1}$ \citep{1995MNRAS.272...41S}, a ratio of stellar luminosity to ionizing photon output $L_*/S_* = 47 \psi_{47}$\,eV,  and a dust opacity to starlight of $10^{-21}  \sigma_{d*,-21}$\,cm$^2$ per H atom \citep{2011ApJ...732..100D}.     

With these choices, ionized gas within the clump can exist in a sequence of states corresponding to increasingly intense irradiation.  The importance of radiation pressure, and the dust optical depth to starlight $\taudstar$ (from the star to the ionization front) both increase along the sequence.  We identify three regimes; see \citet{2011ApJ...732..100D} and Figure 1 of \citet{2012ApJ...757..108Y}.  If the intensity of starlight is sufficiently low, one has a classical Str\"omgren sphere or photo-evaporative flow.  In this state most ionizing photons are absorbed by H atoms; radiation pressure is negligible relative to ionized gas pressure;  and $\taudstar<1$.  For intermediate intensities, radiation pressure is still small; however $\taudstar\simeq 1$ and dust grains absorb a signifcant fraction ionizing photons \citep{1972ApJ...177L..69P}.  For high intensities, H\,II is compressed by radiation pressure into a thin layer and its dynamical influence is relatively weak; dust consumes up to 70\% of the ionizing photons; and $\taudstar$ reaches a maximum of $\taudstarmax=2.0$.  (This value depends weakly with the parameters; see Appendix B of \citealt{2012ApJ...757..108Y} for  details on this radiation-confined limit.) 

We require an estimate for the additional force $F_i$ due to the photo-ionzied gas  (for dynamical mass ejection), as well as its volume-averaged density $n_i$ (for the evaporation rate).   To calculate these we make reference to an idealized, uniform, dust-free H\,II region with same radius as the clump, for which the H density is $[3S/(4\pi \alpha_B \Rc^3)]^{1/2}$. We find that the optical depth across the reference region is $\taudSt = {\sigmac/ \sigmatau}$, where 
\begin{equation} \label{eq:sigma_tau} 
\sigmatau = {0.30 \over T_{i,4}^{0.4}} \left[ (\alphavir/1.6) \psi_{47}  \over (1-f_g) (L/M)_{*,3} \sigma_{d*,-21} \right]^{1/2} \,{\rm km\,s^{-1}}.
\end{equation} 
This indicates that clumps of interest ($\sigmac>1\,\kps$)  will  have optically thick H\,II, except when their stars are relatively dim.  A simple approximation\footnote{For a more precise analysis, see Kim, Kim, \& Ostriker (2015, {ApJ}, submitted).} to the actual dust optical depth of H\,II regions in the \citet{2011ApJ...732..100D} models, valid in both thick and thin limits, is  
\begin{equation} \label{eq:taudstar} 
\taudstar \simeq {\taudstarmax \taudSt\over \taudstarmax +  \taudSt} = {\sigmac\taudstarmax\over\sigmatau\taudstarmax  + \sigmac}.  
\end{equation} 

For a filled H\,II region with radius $\Rc$, the volume-averaged ionized gas density $\nbari$ is related to the optical depth $\taudstar$ by $\taudstar = \nbari \Rc\sigmadstar$.  If the H\,II  is compressed into a thin shell,  $\taudstar = \nbari \Rc\sigmadstar/3$.  An approximation for $\nbari$, which is valid in both thick and thin limits, is therefore 
\begin{equation}\label{eq:nibar}
\sigmadstar \Rc \nbari \simeq  {3\taudstarmax\sigmac\over3\taudstarmax\sigmatau  + \sigmac}. 
\end{equation} 

We can now calculate the net outward force  due to the existence of the H\,II gas, $\FII= 4\pi \Rc^2 \times 2.2 \nbari k_B T_i$.  Computing this, and comparing to the force of starlight, 
\begin{equation} \label{eq:F_i}
{\FII  \over L/c} \simeq {\sigma_F^2 \over \sigmac (\sigmac + 3\taudstarmax \sigmatau)} 
\end{equation} 
where 
\begin{equation} \label{eq:sigma_F} 
\sigma_F  = 2.8\left[ (\alphavir/1.6) (\taudstarmax/2) T_{i,4} \over (1-f_g) (L/M)_{*,3} \sigma_{d*,-21} \right]^{1/2} {\rm km\,s^{-1}}. 
\end{equation} 
 
Equation (\ref{eq:F_i}) agrees with the approximation employed by \citet{2009ApJ...703.1352K} in the limit $\sigmac\ll 3\taudstarmax \sigmatau$, but is more accurate at higher $\sigmac$ because it takes dust absorption into account.  As a result $\FII$ matches $L/c$ for a value of $\sigmac$ which is lower by about a factor of 2.2 in this theory relative to \citeauthor{2009ApJ...703.1352K}'s.  But, since radiation pressure dominates in most of the regime where dust absorption is significant, the improvement is important only for a limited range of clump velocity dispersions (about 1 to 4\,km\,s$^{-1}$). 

A caveat: in equations (\ref{eq:sigma_tau})--(\ref{eq:sigma_F}) we have assumed the H\,II is ionization-bounded with an ionization front radius equal to $\Rc$ -- the state which divides confinement and blowout.   This suffices to estimate the photo-evaporation rate and to identify the criterion for blowout (and perhaps to discriminate compact and extended H\,II regions), but not for any more detailed predictions about the condition of the H\,II gas. 

One might be puzzled that the effects of H\,II gas depend directly on the clump's velocity dispersion $\sigmac$, so we pause to explain.  The first point to note is that the H\,II dust optical depth, and the ratio of its pressure to the radiation pressure, are both functions of a single parameter (which varies along the sequence of states described above).   Second, \citet{2009ApJ...703.1352K} showed that radiation pressure matches H\,II gas pressure at a characteristic radius proportional to $L^2/S$.  For known dimensionless ratios $(L/M)_*$, $(L/S)$, and $f_g$, this corresponds to a radius proportional to $\Mc$, and when $\alphavir$ is known this means a particular value of $\sigmac$.   Therefore, the parameter which controls radiation pressure and dust optical depth in H\,II regions maps onto $\sigmac$, when the comparison is made at the clump radius $\Rc$. 

Note that any evaporative outflow can be affected by gravity if $\vescc$ becomes large compared to the ionized sound speed (10\,km\,s$^{-1}$).  However radiation pressure is strong enough, in this regime, that gravity cannot confine the flow. 

\begin{figure}
\includegraphics[scale=0.4]{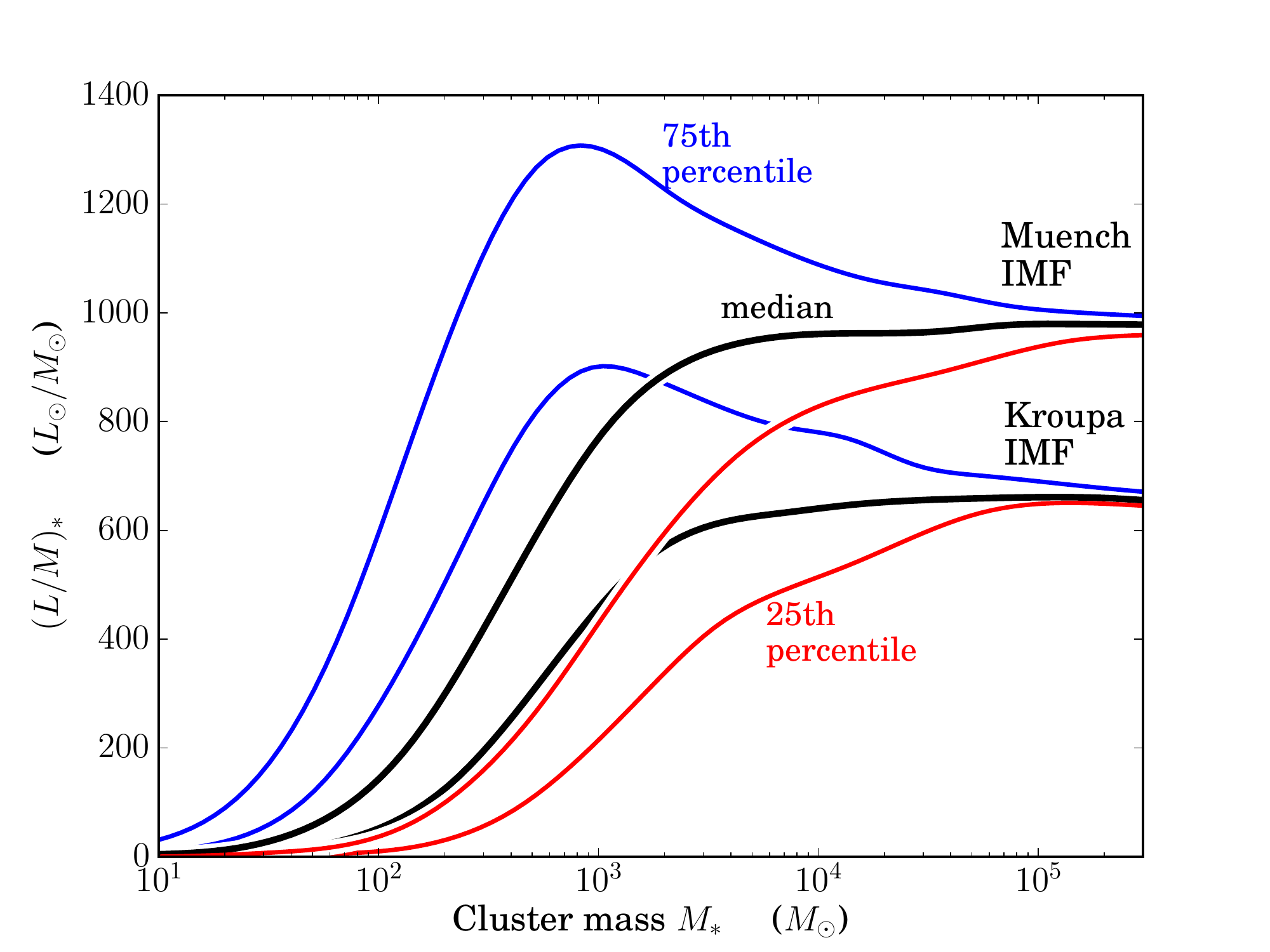}
\caption{Effect of finite sampling on the zero-age stellar luminosity-to-mass ratio, which enters the critical column density $\Sigmacrit$  for radiation pressure feedback (eq.\,\ref{eq:Sigmacrit}). Here stars are drawn randomly from either a \citet{2002ApJ...573..366M} initial mass function, extended with the Salpeter slope to 120\,$M_\odot$, or a \citet{2001MNRAS.322..231K} IMF; clusters are sorted by mass, and $(L/M)_*$ is recorded using the zero-age main sequence fits of  \citet{1996MNRAS.281..257T} at solar metallicity. }
\label{Fig:ClusterSampling}
\end{figure}

\begin{figure}
\includegraphics[scale=0.44]{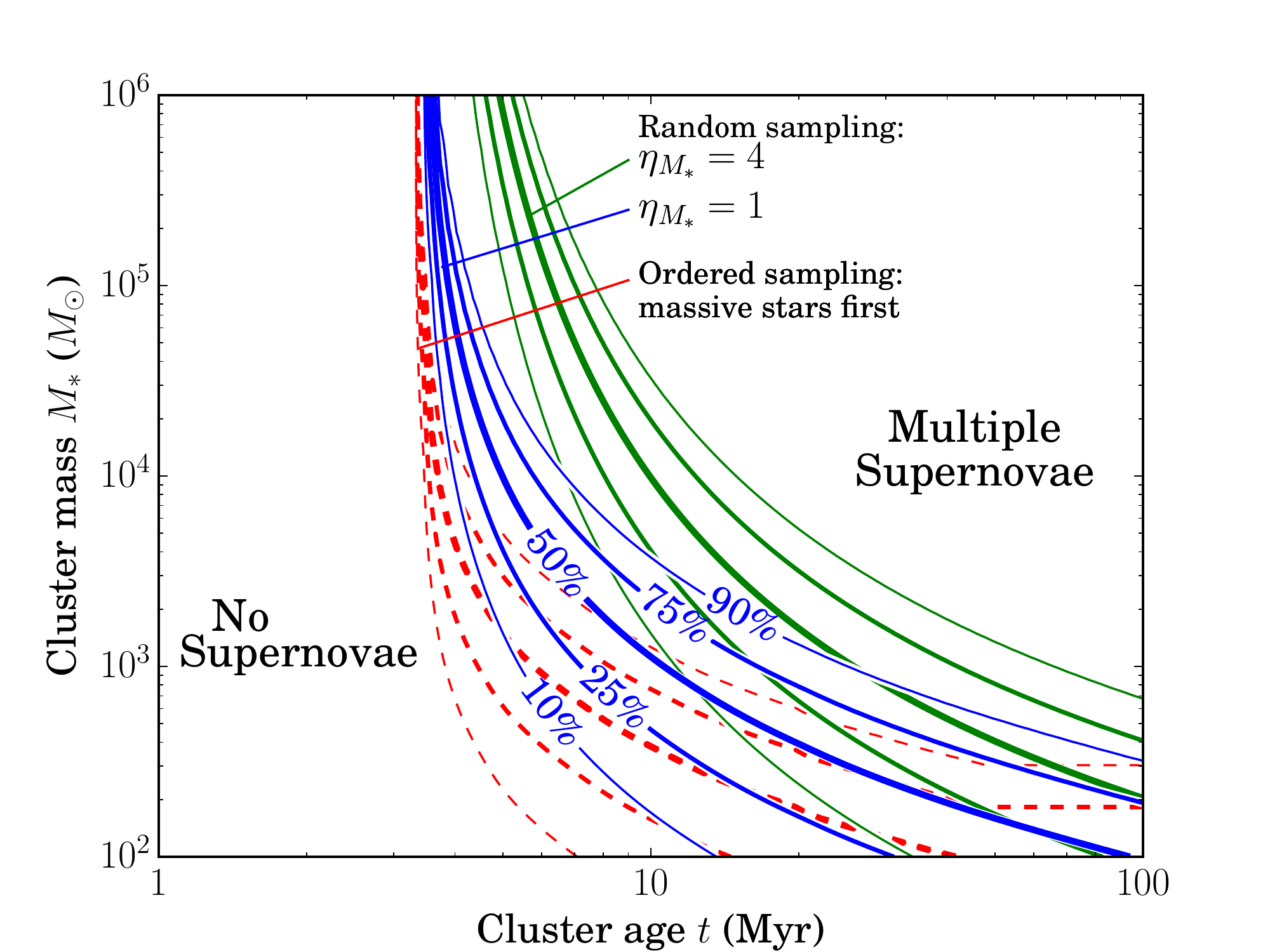}
\caption{Probability that a supernova has exploded at time $t$, if star formation begins at $t=0$ and proceeds as $N_*\propto t^{\eta_{M_*}}$, provided stars are drawn randomly from a \cite{2001MNRAS.322..231K} IMF.  An accelerating star formation law (green lines) contains a younger stellar population at fixed $t$, and hence implies later SNe, than a constant rate of star formation (blue lines).  The probability of a SN is maximized by assuming that the most massive star forms at $t=0$ (red dashed lines).   Each set of lines has identical contour levels.   We use rotating stellar lifetimes from \citet{2013ApJ...764...21C}. }
\label{Fig:SNsampling}
\end{figure}

\subsection{Energy injection and mass ejection by massive stars}\label{SS:massive-ejection} 

Combining the direct radiation, stellar winds, and H\,II gas pressure forces into a net outward force $F_{\rm out}$,
 and comparing against the force of gravity (modified by the indirect radiation force), we estimate 
\begin{eqnarray} \label{eq:Frad+Fion/Fgrav} 
{F_{\rm out} \over F_{\rm grav}' } &\simeq &{\Sigmacrit \over (1-\Gamma )\Sigmac} \left[1+ \phiIR + \phi_w + {\sigma_F^2\over \sigmac(\sigmac + 3\taudstarmax  \sigmatau) }\right] \nonumber \\ &\equiv& {\Sigma_{\rm crit, tot}\over \Sigmac}.
\end{eqnarray} 

We use this combined force to estimate both the creation of turbulent kinetic energy, and the rate of mass ejection, due to massive-star feedback.   For the former, we assume that massive-star feedback accelerates feedback at a rate  $\varphiHMSF F_{\rm out}/M_g$, where $\varphiHMSF$ is a coupling factor. The additional acceleration reduces the external acceleration $\aext$ required to maintain a given $\sigmac$, thereby suppressing $\xic$ according to equation (\ref{eq:xic-versus-phiacc}).  We expect that $\varphiHMSF$ depends on the source of $F_{\rm out}$: for instance, pho-ionization drives turbulence relatively efficiently by the rocket effect, whereas indirect radiation force is very smooth.   In our calculations we employ $\varphiHMSF=0.1$, but we find that the results are not sensitive to this parameter because of the intense mass loss that occurs when $F_{\rm out}$ becomes strong.    

Prior to any supernove, there are two new types of mass loss due to massive stars: dynamical mass ejection and photo-evaporation.  To account for dynamical mass ejection, we posit \citep[e.g.,][]{2010ApJ...710L.142F}  that matter boils away even when $F_{\rm out} < F_{\rm grav}'$, because of inhomogeneities in the  distribution of matter within the clump.   In particular, we assume that all the clump matter with a radial column less than $\Sigma_{\rm crit, tot}$ is blown away each free-fall time, i.e. 
\begin{equation} \label{eq:Blowout} 
{\Mdot_{\rm out, dyn} \tffc' \over M_c} = f_g ~{\cal F}_M\left(\Sigma_{\rm crit, tot} \over\Sigmac\right)
\end{equation} 
where ${\cal F}_{(A,M)}(x)$ is the (area, mass) fraction of the clump with a column below $x \Sigmac$.    

Note that we have assumed that if the forces balance for a gas shell at $\Rc$, it will be blown away.  The forces of starlight and H\,II pressure are continuous and scale similarly to the force of gravity, so we do not believe a dynamical correction would be warranted. 

In addition to dynamical ejection, photo-ionized H\,II gas will escape along open lines of sight, which we estimate to cover a fraction ${\cal F}_A(\Sigma_{\rm crit,tot}/\Sigmac)$ of the clump's surface area.  Multiplying this surface area times the mean ionized gas density $1.4\,m_p \nbari$, and assuming an  outflow speed equal to the ionized sound speed, we derive the mass ejection rate $\Mdot_{\rm evap}$. Comparing this to the clump mass per free-fall time, we find 
\begin{eqnarray} \label{eq:Mdot_evap} 
{\Mdot_{\rm evap} \tffc\over \Mc} &\simeq& 0.41 {(\alphavir/1.6)^{1/2}  (\taudstarmax/2.0) T_{i,4}^{1/2}\over  \sigma_{d*,-21} }  {1\kps \over \Sigma_{c,{\rm cgs}}\sigmac} \nonumber \\ &&~~
\times {{\cal F}_A\left(\Sigma_{\rm crit,tot}\over \Sigmac\right ) \over 1 + 3\taudstarmax\sigmatau /\sigmac} . 
\end{eqnarray} 
For a clump with decreasing $\Sigmac$, photo-evaporation (eq.\,\ref{eq:Mdot_evap}) sets in earlier than dynamical ejection (eq.\,\ref{eq:Blowout}), because blowout affects a larger fraction of the area than of the mass (i.e., ${\cal F}_A(x)>{\cal F}_M(x)$).  However, the characteristic rate of photo-evaporation is several times lower than that of dynamical ejection.  

For calculations, we employ\footnote{Todd Thompson and Mark Krumholz introduced this mass-loss prescription in the context of mass ejection from starburst galaxies (2014, private communication). } a log-normal distribution of column densities, so that
\begin{equation} \label{eq:lognormal-Blowout} 
{\cal F}_{(A,M)}(x) = \frac12\left[1 + {\rm erf}\left( \ln x  \pm \sigma_{\ln x}^2/2 \over \sqrt{2 }\sigma_{\ln x} \right)\right]
\end{equation}
where $\sigma_{\ln x}$ is the standard deviation of $\ln x$.  For the column distributions of nearby molecular clouds \citep{2008A&A...489..143L,2010MNRAS.406.1350F} $\sigma_{\ln x}$  is about 0.4, so we adopt this value.  

\subsection{Combined models: accretion, outflows, and massive stars} \label{SS:Combined-models}

To account for all three effects, we adopt an iterative approach.  We start with an Eddington factor $\Gamma=1$.   Beginning with the model for accretion and outflows worked out in \S \ref{S:Feedback}, we estimate the clump luminosity $L$ assuming that $(L/M)_*$ takes its median value for the stellar mass $M_*$ (figure \ref{Fig:ClusterSampling}).  We work out $\Sigma_{\rm crit,tot}/\Sigmac$ from equation (\ref{eq:Frad+Fion/Fgrav}).  We add the mass ejection from massive stars, from equations (\ref{eq:Blowout}) and (\ref{eq:Mdot_evap}), to that from protostellar outflows, and use this in equation (\ref{eq:fgFromMdotejPerStar}) to determine the self-consistent value of $f_g$.  Likewise we add the turbulent kinetic energy generated by massive-star feedback, and use this to re-calculate $\xic$.  From the values of $f_g$ and $\Sigmac$ we derive the dust temperature $T_d$  from equation (\ref{eq:Td_appx}) and the Eddington factor $\Gamma$ from equation (\ref{eq:GammaEdd_approx}).  To account for indirect force, we replace $G$ with $(1-\Gamma)G$ in any instance where gravity acts on gas, and start over.  Because $\Gamma$ is not large, our solution converges rapidly.    Throughout this process we hold $\alphavir'$ fixed at its fiducial value of {\bff 1.6}, and adopt the \citet{2001MNRAS.322..231K} IMF. 

\begin{figure}
\includegraphics[scale=0.44]{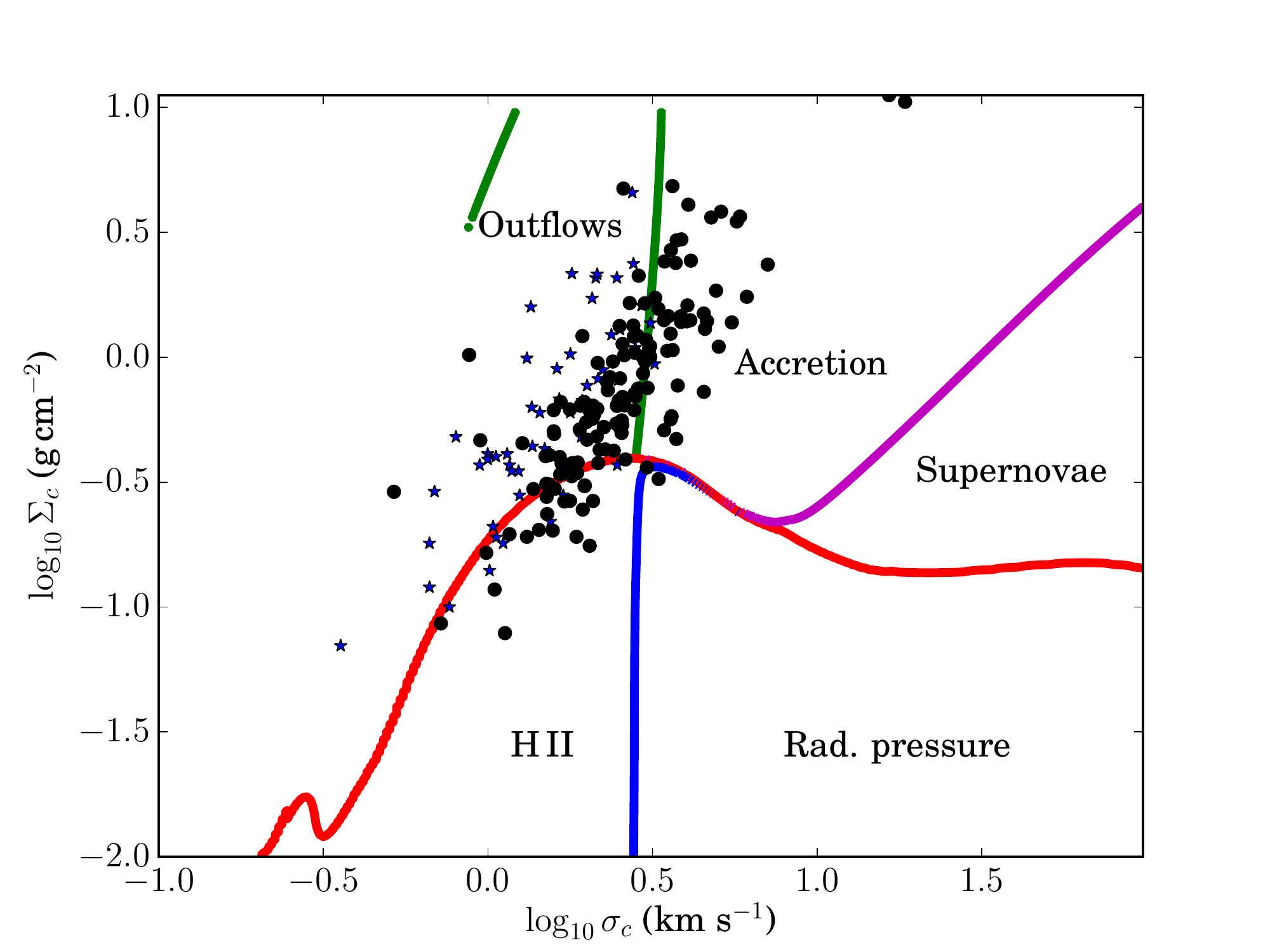}
\caption{Phase space of feedback for accreting clumps and embedded clusters.  Median IMF-sampled values are assumed for the properties of the star cluster.  The `outflows' region corresponds to $>50$\% turbulent energy from outflows; `H\,II' corresponds to $F_{\rm out}>F_{\rm grav}$ and $\FII < F_{\rm out}/2$; `Rad.\ Pressure' corresponds to $F_{\rm out} > F_{\rm grav}$ and $\FII > F_{\rm out}/2$; and `Supernovae' corresponds to a greater than 50\% chance of a supernova explosion (with $\eta_{M_*}=2$ and random sampling; see Figure \ref{Fig:SNsampling}). 
 Massive star forming regions from \citeauthor{2004A&A...426...97F} (\citeyear{2004A&A...426...97F}, {\em points}) and \citeauthor{2003ApJS..149..375S} (\citeyear{2003ApJS..149..375S}, {\em stars}) are plotted for comparison.}
\label{Fig:FeedbackPhases}
\end{figure}

\begin{figure}
\includegraphics[scale=0.44]{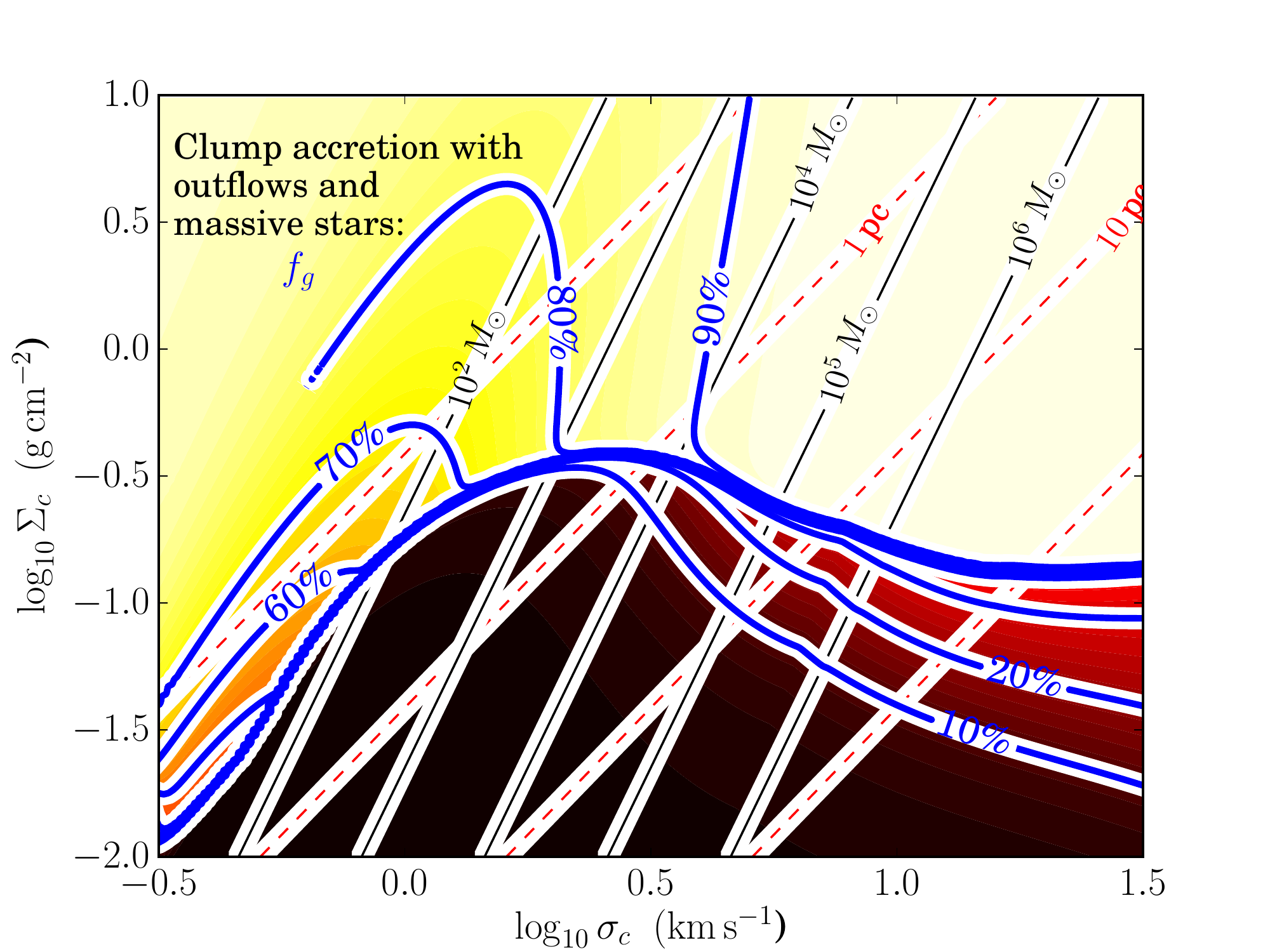}
\caption{Gas mass fraction $f_g$ (color and blue contours) in a model for accreting clumps affected by outflows from low-mass stars as well as ionization, stellar winds, and radiation pressure due to massive stars.  Note that stellar evolution off the main sequence, important in the lower-right region of this plot, is not taken into account. }
\label{Fig:HMSF_fg}
\end{figure}

The outcome of this exercise is plotted in figure \ref{Fig:FeedbackPhases}, in which we outline where in the phase place of growing clusters each effect is dominant, and in figure \ref{Fig:HMSF_fg}, in which we display the gas fraction $f_g$ computed in the manner outlined above.  Our models display a sharp transition from gas-rich to gas-poor as massive stellar feedback becomes important, reflecting the fact that gas disruption accelerates as the gas fraction decreases.  (Because of this rapid transition, we evaluate $q_*$ in equation \ref{eq:fgFromMdotejPerStar} rather than assuming self-similar growth with $M_*\propto M_c$ and $q_*=1$; however this modification does not strongly affect the outcome.)

\section{Comparison to observed regions}\label{S:ObsComparison}

We compare against two types of regions: individual, relatively small regions NGC~1333 and Serpens South, both of which have been studied in detail, and massive regions in the surveys described by \citet{2003ApJS..149..375S} and \citet{2004A&A...426...97F}. 

\subsection{NGC 1333}\label{SS:NGC1333}
The NGC 1333 star forming region within the Perseus molecular cloud provides a point of comparison for theories of outflow feedback.  NGC 1333 
hosts 137 known young stellar objects, of which about 39 are of Class 0 or Class I \citep{2008ApJ...674..336G}.  The population of near-infrared excess sources indicates a typical YSO age 1-2 Myr \citep{1996AJ....111.1964L}.   \citet{2008PASJ...60...37H} identify a distance $D= 235\pm18$\,pc based on maser astrometry, so we adopt this value.   We infer from \citet{2008ApJ...674..336G} that the area of the embedded cluster corresponds to a radius in the vicinity of 0.4\,pc, and identify this with the clump radius $\Rc$.  This is intermediate between the radii of $^{13}$CO and C$^{18}$O(1-0) emission observed by  \citet{2003AJ....126..286R} (0.46\,pc and 0.27\,pc, respectively, at $D=235$\,pc) so we adopt an intermediate column density $\Sigmac=0.19$\,g\,cm$^{-2}$ and velocity dispersion $\sigmac=1.1$\,km\,s$^{-1}$, with uncertainties of about 20\%.    (The \citealt{2003AJ....126..286R} analysis corresponds to $\Sigma = (0.2, 0.17)$\,g\,cm$^{-2}$ and $\sigma=(1.2, 0.85)$\,km\,s$^{-1}$ for the scales probed by the two lines.)  For the adopted values, $\Mc = 460\,M_\odot$ and $\alphavir= 1.1$.   If we assign an average mass of $0.5\,M_\odot$ per YSO \citep{2009ApJS..181..321E},  $f_g\simeq 0.87$.  (We note that \citet{2010ApJ...715.1170A} identify a larger, more diffuse region with NGC\,1333 [$R= 2$\,pc, $\sigma=0.93$\, km\,s$^{-1}$, $\Sigma = 0.02$\,g\,cm$^{-2}$], but since this region has a crossing time of 2\,Myr, we associate it instead with an infall or feeding zone.)

Combining FCRAO and CARMA observations of $^{12}$CO and $^{13}$CO(1-0) emission, \citet{2013ApJ...774...22P} provide the most sensitive analysis  to date of protostellar outflows within NGC 1333, at least within the (0.48\,pc)$^2$ sub-region mapped.  
 \citeauthor{2013ApJ...774...22P}\ identify 22 outflow lobes with typical dynamical ages $\sim5\times 10^4$\,yr and a current, inclination-corrected, total outflow momentum of roughly 35\,$M_\odot\,$km\,s$^{-1}$.   Dividing by the dynamical age leads to a net outflow force of roughly $700\,M_\odot$\,km\,s$^{-1}$\,Myr$^{-1}$.  If this force coupled only to the estimated $\sim 143 M_\odot$ of clump material within region mapped by \citeauthor{2013ApJ...774...22P}, the resulting acceleration of 5\,km\,s$^{-1}$\,Myr$^{-1}$ would exceed the turbulent acceleration $\sigmac^2/\Rc \simeq 3.1$ \,km\,s$^{-1}$\,Myr$^{-1}$.    This comparison is not entirely satisfactory, however, because it involves only a portion of the clump.  
 
 \citet{2010ApJ...715.1170A}, who survey the entire Perseus cloud, provide a more complete view.   In the NGC\,1333 region,  \citet{2010ApJ...715.1170A} infer at least 74\,$M_\odot\,$km\,s$^{-1}$ in known outflows and new outflow candidates (and reckon that this may underestimate the total momentum by as much as a factor of seven).   With a typical dynamical age of $5\times 10^4$\,yr, this corresponds to a net outflow force of at least $1480\,M_\odot$\,km\,s$^{-1}$\,Myr$^{-1}$ and an outflow-driven clump acceleration scale of at least 3.2\,km\,s$^{-1}$\,Myr$^{-1}$.  Outflows are clearly relevant and probably sufficient to drive the observed turbulence. \citeauthor{2010ApJ...715.1170A}\ and \citeauthor{2013ApJ...774...22P}\ come to the same conclusion by comparing outflow, turbulent, and gravitational energies. 
 
The state of NGC 1333 appears to be consistent with the theory of M07 as presented in \S~\ref{S:Feedback}.   Adjusting our parameters to match the observations ($\alphavir = 1.1$, $\SFRff=0.03$, $\vchar = 21$\,km\,s$^{-1}$), we would predict that 80\% of the turbulent energy derives from outflow driving, with accretion providing the remainder.  This accretion could be along the dense molecular filament apparent in \citealt{2012A&A...540A..10S}'s {\em Herschel} map of the region, and another, less prominent filamentary structure extending to the east.   A rough estimate of the column density  of this structure gives about $1.5\times 10^{22}$ H atoms cm$^{-2}$ across an effective width of 0.2\,pc in each filament, for a net mass per unit length $3\lambdafil \simeq 34\,M_\odot$\,pc$^{-1}$. Equation (\ref{eq:Mdot_in_Filamnetary_Infall}) then suggests an infall rate of $\sim 77\,M_\odot$\,Myr$^{-1}$. 

M07, in contrast, found the current outflows insufficient to drive turbulence in NGC 1333.  However M07's analysis was based  on a much lower outflow momentum ($10\,M_\odot\,\kps$, from \citealt{2000A&A...361..671K} and \citealt{2005ApJ...632..941Q}).   An upward revision of outflow momentum by is  perhaps not surprising, considering the corrections for optical depth, finite sensitivity, velocity range, and excitation temperature in observations of protostellar outflows highlighted   \citet{2014ApJ...783...29D},   which amount to nearly an order of magnitude typical increase of momentum and force relative to uncorrected values.  

However, the statistical analyses of the NGC\,1333 region by \citet{2009ApJ...707L.153P} and \citet{2009A&A...504..883B} give reasons for caution.  \citeauthor{2009A&A...504..883B}\ use a principal component decomposition of CO maps of the region, finding no hint of local turbulent driving in the $^{13}$CO  map and only tentative evidence for turbulent driving on 0.4-0.8pc scales in the C$^{18}$O map (which highlights denser matter on scales of the actual clump).  It is unclear whether this result is at odds with our conclusions in \S~\ref{SS:NGC1333}, considering that we invoke a combination of local and large scale (i.e. accretion) driving, and considering that \citet{2010ApJ...722..145C} identify biases in the principal component method when applied to outflow-driven turbulence. 

\citeauthor{2009ApJ...707L.153P}'s analysis of the $^{13}$CO map is more starkly at odds with our conclusion of strong outflow feedback,  as it employs velocity information (via the velocity coordinate spectrum method of \citealt{2006ApJ...652.1348L}) and shows no indication of any deviation from simulations of turbulence driven isotropically on large scales.   This is puzzling, given that the NGC~1333 clump represents a strongly self-gravitating region of the Perseus cloud, which is distinct  from the turbulent background in both column density \citep[e.g.][]{2009A&A...508L..35K} and in the line width-size relation \citep[e.g.][]{1995ApJ...446..665C}.  \citet{2010ApJ...715.1170A} point out that identifying a driving scale is likely to be more difficult by when driving occurs over a range of scales, and indeed this is a critical feature of outflow driving in the M07 theory.  Similarly, \citet{2010ApJ...722..145C} have stressed that driving by collimated outflows leaves a different imprint on turbulent motions than isotropic driving.  Clearly the \citeauthor{2009ApJ...707L.153P}\ result merits further investigation.  

\subsection{Serpens South} \label{SS:SerpensSouth}

The Serpens South cluster-forming region was discovered by \citet{2008ApJ...673L.151G}, who report 
a total of 91 protostars within the cluster boundary, and a high fraction of Class I sources indicative of a very young age (0.1-0.3 Myr).  
The distance is thought to match that of the Serpens main cluster, for which a photometric estimate of 260$\pm$37\,pc \citep{1996BaltA...5..125S} conflicts with a more recent VLBI parallax of 429$\pm$2\,pc \citep{2010ApJ...718..610D}. 
The elongated dust emission implies a gas mass  (420 to 560)$(D/429\,{\rm pc})^2\,M_\odot$ in three concentrations with scale $\sim 0.28\,{\rm pc} (D/429\,{\rm pc})$; the velocity dispersion is 0.3-0.5\,km\,s$^{-1}$ from N$_2$H$^+$ \citep{2013ApJ...778...34T} or  $1.0-1.3$\,km\,s$^{-1}$ in HCO$^+$ \citep{2011ApJ...737...56N}.   With these numbers, $\alphavir\sim 0.8 (429\,{\rm pc}/D)$.   

With its filamentary structure, very young age, and less-than-unity virial parameter \citep{2013ApJ...778...34T}, the Serpens South region resembles the initial conditions for cluster formation (\S~\ref{S:Accumulation}) more than our model for a cluster-forming clump.   Nevertheless, a burst of outflows accompanies the burst of star formation, so we briefly revisit the question of outflow feedback previously addressed by \citet{2011ApJ...737...56N} and \citet{2015ApJ...803...22P}.  These authors observe different transitions ($^{12}$CO(3-2) and $^{12}$CO(1-0), respectively) but agree on a total outflow momentum of (21 to 25)$(D/429\,{\rm pc})^2 M_\odot$\,km\,s$^{-1}$ and outflow dynamical age $\sim 2.4(D/429\,{\rm pc})\times 10^4$\,yr.    This suffices to accelerate the clump matter at a rate $\sim 2$\,km\,s$^{-1}$\,Myr$^{-1}$,  comparable to the turbulent acceleration $\sim 3$\,km\,s$^{-1}$\,Myr$^{-1}$ (both $\propto (D/429\,{\rm pc})^{-1}$).   In addition, there is accretion along and perpendicular to the filament at rates $\sim(50, \,210)(D/429\,{\rm pc}) M_\odot/$yr, respectively \citep{2013ApJ...766..115K}, and this alone suffices to drive $\sigmac\sim 0.8$\,km\,s$^{-1}$ for $\xic\simeq 1$.   We therefore concur with previous analyses that the Serpens South proto-cluster is in its initial stages of growth, but is already affected by outflow feedback. 

\subsection{Massive star forming regions}\label{SS:massiveregions} 

Comparing the parameters of the massive cluster forming regions from \citet{2004A&A...426...97F} and \cite{2003ApJS..149..375S} with the theory outlined here, as in Figure \ref{Fig:FeedbackPhases}, we see that most of these regions have parameters for which protostellar outflows are expected to be an important contribution to the turbulent velocity.  Very few regions are found in conditions for which we would  expect rapid gas dispersal by H\,II or by radiation pressure, while some (in the range 1\,km\,s$^{-1} < \sigmac< 3$\,km\,s$^{-1}$) are near the boundary of this regime.    

However, a striking feature of the selected population is the correlation between $\sigmac$ and $\Sigmac$, which roughly corresponds to $\Rc\propto \Mc^{1/4}$ and $\Sigmac\propto\sigmac^{4/3}$.  Regions with high mass and high $\sigmac$ are found at high column densities, well above the critical column at which radiation pressure becomes important.   We interpret this to mean that stellar feedback is not significant in these regions, and that their properties are a consequence of the initial conditions for cluster formation rather than the action of their stars.

{\bff 
\section{Comparison to previous models}\label{S:Discussion-prevmodels} 

\citet{1999PhDT........11M}, \citet{1999sf99.proc..353M}, \citet{2006ApJ...644..355H,2007ApJ...666..281H}, \citet{2010ApJ...710L.142F},
 \citet{2012ApJ...751...77Z}, and \citet{2014ApJ...793...84Z} all present analytical star cluster formation models that overlap ours in some respects, so a comparison is warranted.    

\citeauthor{1999PhDT........11M}, considering only protostellar outflow feedback, evaluates mass ejection using the MM00 models and accounts for the generation of turbulence by outflow impulses.   His models neglected clump accretion and massive-star feedback, but include a treatment of the clump's energy equation (which we replace with the cruder assumption of a constant virial parameter $\alphavir$).  They display the feedback-driven oscillations discussed in \S~\ref{SS:outflows-energetics-stability}.  

\citeauthor{2007ApJ...666..281H} solve the clump energy equation within an extended clump, assumed to be instantaneously isothermal but with an evolving effective sound speed.  As they  neglect feedback and clump accretion, their clumps contract over a couple free-fall times  toward an unstable state of high central concentration, while forming stars at an accelerating rate. 

\citeauthor{2010ApJ...710L.142F} consider the same set of feedback phenomena we consider here, but concentrate on the conditions under which a high stellar mass fraction leads to rapid disruption of the remaining gas.   Although their analysis is very similar to ours (except for details like dust opacity of H\,II, for instance), \citeauthor{2010ApJ...710L.142F} did not consider the evolution of growing star clusters.  \citeauthor{2010ApJ...710L.142F}  do not address how a clump might produce massive stars (requiring high column densities) and only later blow away gas (requiring low $\Sigmac$).  We have found that accretion provides a natural explanation, at least for the mass range in which stellar feedback appears to be important. 

\citeauthor{2014ApJ...793...84Z} consider star cluster formation as the end phase of a freely-falling cloud in which star formation proceeds as local regions cross a critical density threshold. \citeauthor{2014ApJ...793...84Z} account for photo-evaporative mass loss\footnote{We note that \citet{2014ApJ...793...84Z} employ the the photo-evaporation rate of \citet{1994ApJ...436..795F}, which incorrectly associates swept-up mass with ionized  mass; see the discussion above  \citeauthor{1994ApJ...436..795F}'s equation (8) and compare  their equation (11) with \citet{2002ApJ...566..302M}'s equation (19).} 
on a star-by-star basis, but neglect the dynamical recoil due to photo-evaporation \citep{2002ApJ...566..302M,2006ApJ...653..361K,2011ApJ...738..101G} as well as the other agents of feedback.  Similarly, \citeauthor{2014ApJ...793...84Z} account for mass gain in the form of accretion from the galactic disk, but  neglect the driving of turbulence within the collapsing cloud due to the arrival of fresh material.   They find a runaway acceleration of star formation in low-mass clouds ($\lesssim 10^{4.5}M_\odot$), which is prevented by mass loss in more massive clouds. 

A distinctive feature of the current work is that we separate the virialized cluster-forming clump from its collapsing reservoir region, and account (in approximate ways) for the driving of turbulence both by accretion and by several forms of stellar feedback.    Clumps in our model grow in a roughly self-similar fashion, evolving to larger radii and lower column densities, and becoming increasingly susceptible to massive-star feedback, as they gain mass.  Importantly, our model also predicts an accelerating star formation rate so long as the reservoir supplies an accelerating mass accretion rate, because the rates are nearly proportional  so long as $\epsin$ and $\SFRff$ change slowly.  }

\section{Conclusions} \label{S:conclusions}

In the effort to understand the interaction between gas accretion and stellar feedback, we have developed an approximate model for the growth of a cluster-forming clump by accretion from its environment (\S~\ref{S:Accumulation} and \S~\ref{S:Params}) and the  effects of protostellar outflows (\S~\ref{S:Feedback}) and massive stars (\S~\ref{S:Radpressure}) on the clump's evolution.   In the presence of ongoing accretion, protostellar outflows are expected to be important for relatively low clump velocity dispersions ($\sigmac\lesssim 3$\,km\,s$^{-1}$), and remove matter only gradually.   In contrast, radiation pressure affects massive regions with low column densities ($\Sigmac\lesssim 0.3$\,g\,cm$^{-3}$) and causes rapid gas dispersal.  Photo-ionization and photo-evaporation enhance the disruptive effects of massive stars in a limited range of velocity dispersions (1\,km\,s$^{-1}\lesssim\sigmac\lesssim 4$\,km\,s$^{-1}$).   The well-studied region NGC~1333 appears to be consistent with this analysis, as its complement of protostellar outflows is sufficient to accelerate turbulent motions at nearly the rate required to sustain them (with accretion powering the remainder).  Indeed many of the massive regions in the surveys by \citet{2004A&A...426...97F} and \citet{2003ApJS..149..375S} are expected to be in this state.  Many of these regions are close to the condition\footnote{We stress that this condition is a stochastic one, because the cluster's ionization rate depends on its most massive stars, and because the IMF is not completely sampled in regions for which photo-ionization is important (Figure \ref{Fig:ClusterSampling}).   Our evaluation in Figures \ref{Fig:FeedbackPhases} and \ref{Fig:HMSF_fg} used the median $L$ and $S$ at each $M_*$.} for disruption by photo-ionization, but few exist where we would expect gas to already have been cleared.   

With these results in mind we return to the question of how star cluster formation terminates, because the efficiency and rapidity of this event is crucial to the production of a bound system, and because the birth radius of such a system is crucial to its long-term survival \citep{2011MNRAS.411.1258P}.     The combination of accretion and stellar feedback provides one hypothetical scenario, in which the clump column density $\Sigmac$ declines over time while the complement of massive stars grows, until the conditions for disruption by massive-star feedback are met.    Accretion is an essential element, because the incorporation of fresh material causes $\Sigmac$ to decline (unless the accretion rate accelerates sharply; see \S~\ref{S:Accumulation}).   However star cluster birth could end otherwise, if the mass reservoir is exhausted before feedback plays any role.   

{\bff The feedback-termination scenario is likely to be characterized by rapid gas los.  Given that the gas mass fraction remains high prior to gas expulsion (Figure \ref{Fig:HMSF_fg}), rapid gas loss is likely to unbind the stellar population,} or at least cause it to expand significantly -- as  \citet{1983ApJ...267L..97M} inferred from the radii of embedded and open star clusters.   Mass exhaustion, by contrast, would efficiently form bound clusters, unless the initial conditions are very different from our expectation (\S~\ref{S:Accumulation}) of quiescent infall. 

Another difference  involves the conditions in which most cluster-forming environments are found.  If stellar feedback ends star cluster formation, then regions should be found close to the threshold for gas disruption (unless this is precluded by the selection of regions).   In the surveys of \citet{2004A&A...426...97F} and \citet{2003ApJS..149..375S}, plotted in Figure \ref{Fig:FeedbackPhases}, low-dispersion regions ($\sigmac\lesssim 3$\,km\,s$^{-1}$, which represents most of their data) appear to be consistent with feedback termination.  However these surveys extend for higher $\sigmac$, to column densities well above our estimate of the disruption threshold.   Unless this is due to a selection effect, we conclude that reservoir exhaustion, rather than stellar feedback, ends the formation of the most massive clusters in these surveys.  

We have simplified our analysis by making several key assumptions, all of which require further scrutiny.  We assume that one can divide the cluster-forming `clump' from its accretion flow,  and that the clump will evolve through a series of equilibrium states so long as accretion persists.   The possibility of energetic instability (\S~\ref{SS:Accretion-energetics-stability} and \S~\ref{SS:outflows-energetics-stability}) gives reasons for caution.   Moreover, we have assumed that the evolution can be approximated with a fixed virial parameter $\alphavir$, rather than solving the energy equation directly.  

Our analysis relies on previous work, especially our own papers on stellar feedback (e.g., MM00, M07,  \citealt{2009ApJ...703.1352K}, \citealt{2010ApJ...710L.142F}, and  \citealt{2011ApJ...738..101G},) so we end by noting which elements are new.   One new element, which is also our motivation, is that we incorporate simultaneous infall and feedback within analytical models for star cluster formation.  Another is the introduction of the dimensionless parameter $\xic$ (equation \ref{eq:xic}), which helps to make this combination analytically tractable.   A third is the observation that the mass ejection by outflows can be expressed analytically (equation~\ref{eq:MdotejOnMdotstarFromM07}).  A fourth is our treatment of radiation pressure feedback, which employs an approximate Eddington factor for the indirect force and provides an estimate of the dust temperature.   A fifth is our model for photo-ionization feedback (\S~\ref{SS:Photoionization}), in which we account for the effects of dust opacity in the ionized gas, and for both photo-evaporation and dynamical disruption (albeit in an approximate way).

\acknowledgements
CDM and PHJ are supported by a Discovery Grant from NSERC, the Natural Sciences and Engineering Research Council of Canada.  PHJ is also supported by a Connaught fellowship from the University of Toronto.  We thank Rachel Friesen, Alyssa Goodman, Patrick Hennebelle, Peter Martin, Christopher McKee, Norm Murray, Phil Myers, Quang Nguy$\tilde{\hat{\rm e}}$n L'o'ng, Eve Ostriker, Adele Plunkett, Ralph Pudritz, and Aaron Skinner for helpful and stimulating discussions.  We are especially grateful for comments from our referee, Robi Banerjee, that stimulated us to significantly improve this work. 

\bibliographystyle{apj}
\bibliography{StarFormation}

\end{document}